\documentclass{aastex}
\usepackage{emulateapj5,epsfig}
\newcommand{\etal}{{et~al.}}

\newcommand{\lx}{ergs s$^{-1}$}
\newcommand{\lxh}{$h_{50}^{-2}$ ergs s$^{-1}$}
\newcommand{\bggmph}{($h_{50}^{-1}$~Mpc)$^{1.77}$}
\newcommand{\bggmp}{Mpc$^{1.77}$}
\newcommand{\eband}{$0.3-3.5$~keV}
\newcommand{\ebandp}{($0.3-3.5$~keV)}
\newcommand{\hkpc}{$h_{50}^{-1}$~kpc}
\newcommand{\hmpc}{$h_{50}^{-1}$~Mpc}
\newcommand{\rosat}{{\it ROSAT}}

\newcommand{\figdirthree}{}

\newcommand{\myfigure}{figure*}

\newcommand{\rscale}{1.55}
\newcommand{\figscale}{1.0}
\newcommand{\rheight}{0.4}

\newcommand{\cmheight}{0.2}
\newcommand{\myacknowledgments}{\acknowledgments}

\submitted{Received 2001 June 1; accepted 2001 October 4}
\journalinfo{The Astrophysical Journal}
\shorttitle{New X-ray Clusters in the EMSS II}
\shortauthors{Lewis \etal}

\begin{document}

\title{New X-ray Clusters in the {\it Einstein} Extended
Medium Sensitivity Survey II: Optical Properties}

\author{Aaron D. Lewis\altaffilmark{1,2}, E. Ellingson\altaffilmark{1}, and
John T. Stocke\altaffilmark{1}}

\email{lewisa@uci.edu,
e.elling@casa.colorado.edu, stocke@casa.colorado.edu}

\affil{Center for Astrophysics and Space Astronomy, University of
Colorado, 389 UCB, Boulder, CO 80309}
\altaffiltext{1}{Visiting Astronomer, Kitt Peak National Observatory, National Optical Astronomy
Observatories, which is operated by the Association of Universities for Research in Astronomy, Inc.
(AURA) under cooperative agreement with the National Science Foundation.}
\altaffiltext{2}{Current Address: University of California, Irvine, Department of Physics and Astronomy,
4171 Frederick Reines Hall, Irvine, CA, 92697-4575}

\received{}

\accepted{}

\journalid{}{}

\articleid{}{}

\begin{abstract}  
We present optical images for 9 new clusters of galaxies we have found in a reanalysis of the {\it
Einstein} IPC images comprising the Extended Medium Sensitivity Survey (EMSS). Based on the presence of
a red sequence of galaxies in a color-magnitude (CM) diagram, a redshift is estimated for each cluster.
Galaxy overdensities (cluster richnesses) are measured in each field using the B$_{gc}$ statistic which
allows their plausible identification with the X-ray emission.
The nature of our X-ray detection algorithm suggests that most of these clusters have low X-ray surface
brightness (LSB) compared to the previously known EMSS clusters. 
We compare
the optical and X-ray observations of these clusters with the well-studied Canadian Network for
Observational Cosmology (CNOC) subsample of the EMSS, and conclude that the new clusters
exhibit a similar range
of optical richnesses, X-ray luminosities, and, somewhat surprisingly, galaxy populations as the
predominantly rich, relaxed EMSS/CNOC clusters.
\end{abstract}


\section{Introduction\label{sec_intro3}}

X-ray selected samples of clusters of galaxies are one of the most effective tools used to constrain
cosmological models, complementing studies of the cosmic microwave background, and Type 1a suernovae
\citep{bah99}. The cluster X-ray luminosity function is the most accessible observable obtained from a
large, flux-limited, X-ray cluster catalog
\citep[e.g.,][]{vik98b,nic99,rei99,gio01}, and several techniques
have been used to constrain cosmological parameters such as
$\sigma_8$, and $\Omega_{matter}$ \citep[e.g.,][]{don99a,bah98,hen00,voi00}. The {\it Einstein} Extended
Medium Sensitivity Survey (EMSS) sample of X-ray clusters is of special cosmological interest due its
very large sky coverage and moderate depth and sensitivity, which allow it to probe the
very bright end of the X-ray luminosity function (XLF), at $L_X > 5 \times 10^{44}$~\lx~\ebandp, with
detections of clusters at
$z>0.5$, where the leverage for constraining a given cosmology is greater
\citep{don98,don99a,rei99,gio01}.

X-ray-selected samples of clusters have been preferred to optically-selected samples due to the
significantly lower confusion and projection effects in the former method
(see \citealt{gla00} and \citealt{don01} for excellent discussions of optical vs. X-ray cluster
selection) and because X-ray selection provides a simple sample definition using flux limits. However,
in the first paper in this series
\citet[Paper 1 hereafter]{P2_note1}, we have shown that the original EMSS sample exhibited a modest
selection bias against very extended sources, and is complete at the
$72-83\%$ level because it is actually surface brightness limited, not flux-limited. We have described
our re-analysis of the IPC imaging data from which the EMSS sample was drawn in Paper 1. In a combined
effort of database and literature searches, as well as optical imaging, at least 17 clusters were found
to be missing from the EMSS, with an additional 8 clusters statistically predicted to be 
found in a complete optical search.
All of the new clusters are expected to be of a
more diffuse nature, i.e., although they had sufficient total flux to be included in the EMSS,
their X-ray emission was not centrally concentrated enough to be detected by the original EMSS
detection algorithm. Only through the use of larger detection apertures do these new clusters
obtain S/N $>4$. The addition of these new clusters has reduced the previously reported ``negative''
evolution at high redshift ($z>0.3$) and high luminosity ($L_X > 10^{45}$~\lx) in the XLF to only
$1\sigma$ (Paper 1). Within the EMSS, these objects therefore represent an important class of new
clusters, which are possibly low X-ray surface brightness (LSB) clusters. 

The existence of LSB clusters could have an important impact on our understanding of cluster properties.
The original EMSS cluster sample is apparently dominated by virialized, rich systems with well defined
$L_X - T_X - \sigma_v$ correlations \citep[e.g.,][]{mus97,L99}.
Such clusters are more regular and more massive than
optically selected clusters such as those found by other groups (e.g., MORPHS, \citealt{dre99};
\citealt{sma97}; PDCS, \citealt{pos96}; \citealt{zar97}).
The EMSS preferentially selected clusters with a high central surface brightness, which
results in the exclusion of some systems that may be in a disturbed or unrelaxed dynamical state. 
If LSB clusters have been preferentially missed by X-ray surveys, we may not have observed the full
range of properties in X-ray luminous clusters. For example, a low X-ray surface brightness likely
indicates a non-centrally concentrated gravitational potential for the cluster, which could be the
result of a merger in progress, or an otherwise disturbed or non-virialized system resulting in a lower
intra-cluster medium (ICM) density, lower temperature and thus lower X-ray luminosity compared to more
relaxed systems which define the
$L_X-T_X$ relationship
\citep{mus97}. Such clusters
would exhibit a wider range of $L_X$, $T_X$, and $\sigma_v$ values at the same masses, providing
additional clues to the formation and evolution of clusters of galaxies. 
All current X-ray surveys are surface-brightness limited at some level
\citep[even those surveys designed to detect X-ray extent; see][]{ada00}, and we must understand their
selection functions to properly interpret cosmological constraints derived therein. We therefore
present the new clusters we have found missing from the EMSS as potential LSB clusters, and interesting
objects for optical and X-ray study and comparison with already known X-ray clusters.

In Paper 1, we introduced the X-ray detection algorithm and source catalog, and described the effect
the discovery of additional clusters in the EMSS sample had on the cluster X-ray luminosity function. In
this paper we present the detailed results of our optical imaging campaign, which resulted in the
identification of nine of the 17 new clusters of galaxies. In
\S 2 we present the new clusters, describe the methodology used to identify clusters of galaxies in the
sample, and estimate their redshift and X-ray luminosity. In \S 3 we discuss the cluster properties,
comparing them with other well-studied clusters.
We present a summary in \S 4. We have assumed H$_0=50h_{50}^{-1}$~km~s$^{-1}$~Mpc$^{-1}$, $\Omega_0=1$,
and $\Lambda=0$ in our calculations, but our conclusions are not unique to these choices.
 
\section{The Optical Imaging Campaign\label{sec_optical}}

The new clusters presented in this paper are the optical counterparts to very extended X-ray sources in
our catalog. The creation and investigation of the catalog of new IPC detections are described in detail
in Paper 1. In that work, we describe literature and database searches which reveal previous
identifications for some of the 772 catalog sources at various wavelengths; however, the
vast majority of fields had no definitive identification beyond {\bf not} being clusters of galaxies. To
search for clusters among the unidentified sources, we created 3 subsamples (one random, and two
designed specifically to pre-select clusters; see Paper 1 for details of the X-ray selection criteria)
for which we obtained deep multi-color images of 62 fields at the KPNO 2.1m telescope.  In this section
we discuss the clusters of galaxies we have discovered thus far.

Our campaign was designed to detect any grouping of elliptical galaxies following a linear sequence in
color and magnitude, defined as the cluster red sequence (CRS), taken to be the optimum signature of a
cluster of galaxies in the optical. This method is very similar to that described in detail by
\citet{gla00}. Using the T1KA and T2KA direct imaging cameras at the 2.1m telescope, we initially
imaged each field in Gunn $r$ or Kron-Cousins $R$ to identify any galaxy over-density. The field of view
was $5 \times 5\arcmin$ and $10\times 10\arcmin$ for the T1KA and T2KA chips, respectively. Both
cameras obtain a pixel scale of $0 \farcs 305$. We encountered a wide range of seeing from $0 \farcs 8 -
2 \farcs 5$. In order to detect distant clusters to redshifts similar to the highest redshift EMSS
clusters ($z \sim 0.8$), we required $5\sigma$ limiting magnitudes $\sim 24$ in $r$ or $R$ (which
corresponds to M$^*$+2 at $z\sim0.8$). Thus each field was observed for approximately 1800 seconds in
$r$ or $R$. While examining the red image for galaxies, each field was observed for 600 seconds in
Johnson $B$ to detect any bright blue objects which might be QSOs contributing to the X-ray emission. 
If there was a galaxy overdensity present, we further imaged the field in Gunn $g$ or Johnson $V$,
and/or in Gunn
$i$ or Kron-Cousins $I$.
Typical limiting magnitudes (the magnitude at which a star would be detected at S/N~$\geq5$) of
$r=23.5$ and $R=24$ were obtained. Photometric standard stars were observed 3--5 times per night in each
filter, at a range of airmasses, resulting in a photometric accuracy of $\leq0.1$~mag.

The imaging campaign was conducted over three semesters, in December 1997 (T1KA, 4 nights, poor
weather), December 1998 (T1KA, 6 nights, good weather), and May 2000 (T2KA, 6 nights, good weather). We
imaged a total of 27 randomly selected sources, and 35 pre-selected sources (see Paper 1, Tables 6, 8,
\& 10 for the complete list of observed sources). Table
\ref{tab_obs_data} lists details of the observations of the nine clusters we identified in our
observations. It lists by column: (1) catalog number of the source; (2) \& (3) right ascension and
declination of the source, epoch J2000; (3) filter; (4) exposure time in seconds; (5) the 5-$\sigma$
magnitude limit reached in the observation; (6) date of the observation; (7) comments on the
observation. The full source catalog is available from the first author upon request.

\subsection{Data Reduction and Analysis\label{subsec_data}}

After standard image reduction with IRAF, the identification and classification of all objects in each
field and the calculation of an instrumental magnitude for each object were performed using the Picture
Processing Package (PPP) software developed by
\citet{yee91}. This photometry package produces galaxy ``total'' magnitudes (see \citealt{yee91} for
justification),  corrected for overlapping with neighboring objects, which simulations have found to be
accurate to within 2--3\% systematic error. Random uncertainties for individual galaxies near the
completeness limit are
$\sim$ 0.05-0.1 mag. The completeness limit is the magnitude limit above which we expect to detect
$100\%$ of the high surface brightness galaxies present in the field;
it is approximately
$0.5-0.8$~mag fainter than the point source limiting magnitude \citep[see][]{yee91}. We do
detect lower surface brightness galaxies as faint as $\sim25$ $r$-band mag arcsec$^{-2}$, but we do not
expect to be complete for such objects. Star-galaxy separation was performed using an analysis of object
radial profiles. Bright, non-saturated stars were used to measure the PSF, and subsequently all objects
in the field detected at S/N $>2$ were classified by extent as stars or galaxies.  After a correction
for Galactic extinction assuming
$R_V = 3.1$, scaling to the neutral hydrogen column given by W3nH\footnote{Neutral hydrogen data is
from \cite{dic90}. W3nH is a Web version of the nH FTOOL. nH was developed by Lorella Angelini at the
HEASARC. It is a service of the Laboratory for High Energy Astrophysics (LHEA) at NASA/GSFC and the
High Energy Astrophysics Division of the Smithsonian Astrophysical Observatory (SAO). }, photometric
calibration from the standard stars was used to calculate apparent magnitudes and colors, and we then
used a color-magnitude (CM) diagram to detect a CRS. A CRS approximately following the slope and
zeropoint of empirical galaxy color models
\citep{kod97,kod98}, provides a redshift estimate accurate to $10\%$ \citep{gla00}. We then
calculated the apparent over-density of galaxies relative to the field at that redshift using the
B$_{gc}$ statistic. Employing alternative cosmologies (e.g., the currently favored
H$_0=70$~km~s$^{-1}$~Mpc$^{-1}$, $\Omega_0=0.3$, and $\Lambda=0.7$ cosmology) does not change our
photometric redshift estimates by more than $\Delta z = 0.02$.

The galaxy-cluster spatial covariance function \citep{lon79}, quantified by the B$_{gc}$ parameter, is
described in detail in 
\citet{yee99}. Briefly, it is a quantification of the richness of
the clustering environment around a given object. Given a cosmological model, an assumed galaxy
luminosity function and its evolution, and measured mean background galaxy counts, this parameter
reflects the galaxy overdensity around a given object, correcting for the expected spatial and
luminosity distributions of field galaxies and of the associated cluster galaxies at the redshift
of the object. All galaxies within 500~kpc of a central object (generally the brightest cluster
galaxy (BCG) identified from the CM diagram) brighter than a specific magnitude are counted. 
Then the expected number of background galaxies in that area down to the same magnitude is
subtracted.  The number of excess galaxies is normalized to an evolved galaxy luminosity function
at the redshift of the object and then converted to B$_{gc}$ assuming a form for the spatial
distribution of galaxies. 

The specific magnitude limit for galaxy counting is determined for each field individually and is taken
to be the brighter of the completeness magnitude or M$^*_{r/R}$ + 2.5. Background galaxy counts are
those of
\citet*{yee86} and \citet{yee96}. The evolving galaxy luminosity function used to normalize the excess
galaxies is that of \citet{ell91}, which includes  moderate galaxy evolution of M$_*$($z$) $\sim z$
\citep[see also][]{lin99}. Finally, the distribution of the excess galaxies is assumed to be the
standard cluster galaxy power law,
$r^{-\gamma}$ where $\gamma = 1.77$ \citep{sel78}. The total uncertainty in B$_{gc}$ is given by
$\Delta$B$_{gc}/$B$_{gc} = (N_{net} + 1.3^2N_{b})^{1/2}/N_{net}$
\citep{yee99} where $N_{b}$ is the expected number of background galaxy counts and $N_{net} = N_{total}
- N_{b}$ is the number of excess galaxy counts. The factor of 1.3 is an empirical value that accounts
for the clustering of the background galaxies, which causes the error in the background galaxy counts to
deviate from a Poisson error.  This factor is discussed in detail in
\citet{yee86}. For those more familiar with the cluster richness classification of
\citet{abe58}: B$_{gc}$ values of 600 $\pm$ 200, 1000 $\pm$ 200, 1400 $\pm$ 200, 1800 $\pm$ 200, 2200
$\pm$ 200, and $> 2400$ \bggmp{} are comparable to Abell richness classes (ARC) 0, 1, 2, 3, 4, and
5 respectively \citep{yee99}.

Our overall strategy for identifying clusters is similar to that described by  \citet{gla00}. Our
criteria for a cluster candidate was the detection of a CRS, combined with at least a modest spatial
concentration of the red galaxies.
The B$_{gc}$ value was then calculated at the redshift of the CRS and the location of the BCG for the
candidate cluster. To identify a field as a cluster of galaxies, we required an identifiable CRS
combined with a B$_{gc}$ value
$\gtrsim300$~\bggmp, and consistency with the available X-ray data. The range of B$_{gc}$ values
quoted in this paper for a given cluster are found by calculating B$_{gc}$ at the highest and lowest
redshifts estimated for the CRS in the CM diagram.
The B$_{gc}$ value can be used to estimate an expected X-ray
luminosity (and hence, expected flux) via the observed correlation with $L_X$ \citep{yee01},
which we compare to our
observed X-ray fluxes in the four {\it Einstein} IPC apertures used to generate the catalog.
The IPC apertures are circular, with diameters 2.5, 4.7, 8.4, \& 12.2\arcmin; we refer to them
as apertures 1, 2, 3, \& 4, respectively (see Paper 1, as well as the individual discussion of sources
below).
We caution, however, that there is significant scatter in the B$_{gc} - L_X$ relation
\citep{yee01}, so that the $L_X$ values estimated by this process are indicative, not definitive. Unless
otherwise stated, all X-ray fluxes are in the
\eband~energy band, unabsorbed, (corrected for absorption assuming Galactic neutral hydrogen column
density n$_H$ given by W3nH). Luminosities are K-corrected assuming a power-law spectrum with photon
index
$\Gamma=1.5$ ($\alpha=0.5$, following
\citealt{hen92_h92}; consistent with a cluster spectrum in this bandpass) and quoted in the
\eband~energy band in the rest frame of the cluster.

\subsection{New Clusters in the EMSS\label{subsec_newclusters}}

In Table \ref{tab_newclusters3}, we present a summary of the X-ray and optical properties of the nine
clusters we have discovered optically, which includes by column: (1) catalog source number;
(2) estimated redshift range from photometry; (3) measured galaxy over-density, B$_{gc}$ in
\bggmph; (4) log of the X-ray luminosity in ergs s$^{-1}$ in the
\eband~IPC bandpass calculated from the third IPC aperture flux\footnote{
We have calculated the X-ray luminosity in our given cosmology to more easily compare to previously
published work on EMSS clusters.};
(5) the blue galaxy fraction (see \S~\ref{sec_discussionp3}); (6) ratio of detected X-ray flux in IPC
aperture 3 to aperture 1, indicating extent (see \S~\ref{sec_discussionp3});
(7) notes on the nature or identification for
the cluster. For columns (2), (3), and (4), the best-fit value from the range is given in square
brackets following the range. In this section we discuss details of the individual clusters. The other 8
new clusters in our X-ray catalog were found by other groups conducting X-ray cluster surveys (see
Paper 1).

\begin{figure*}[t]
\parbox{0.49\textwidth}{
\centerline{\psfig{figure=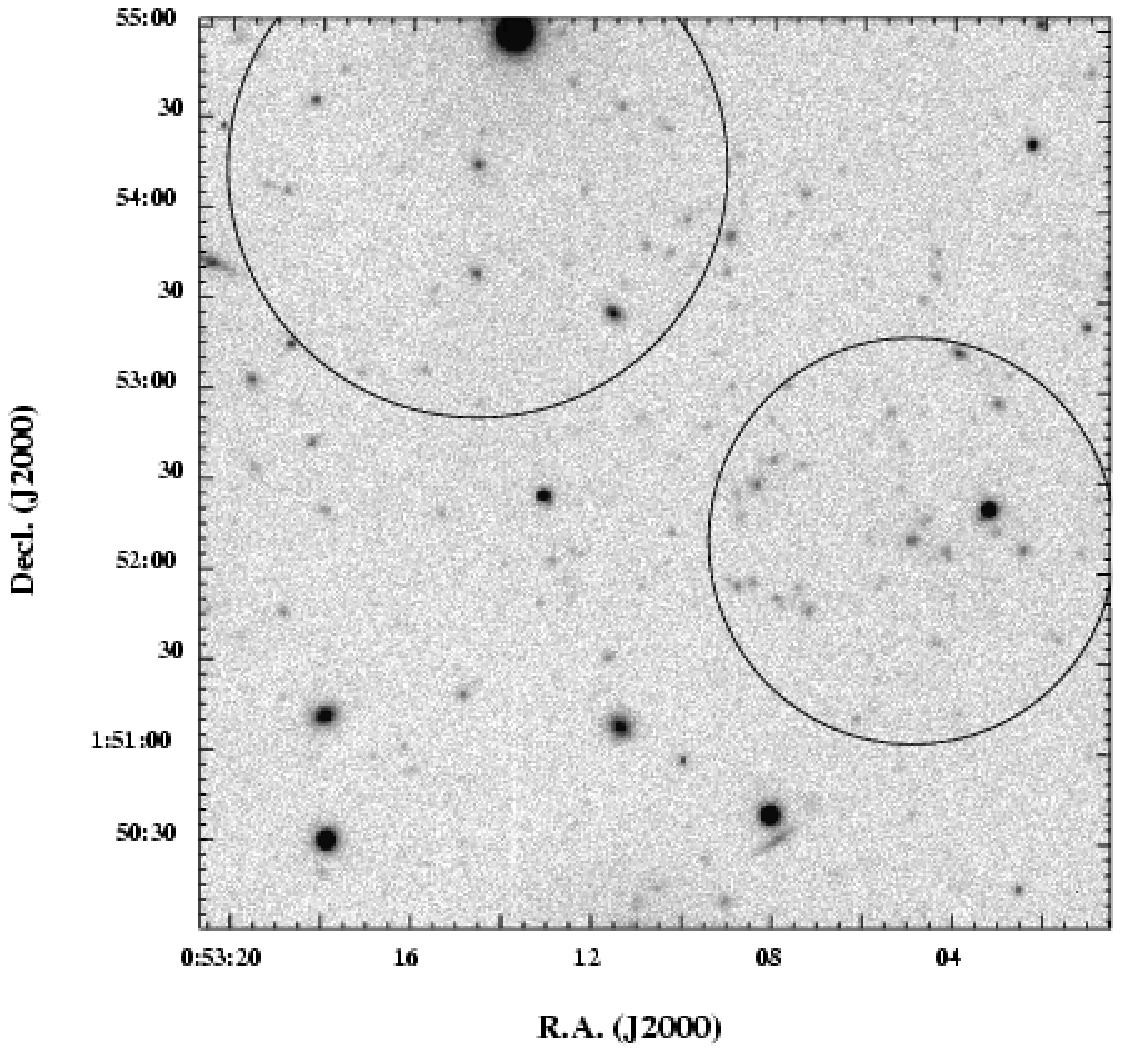,height=\rheight\textheight}}}
\parbox{0.49\textwidth}{
\centerline{\psfig{figure=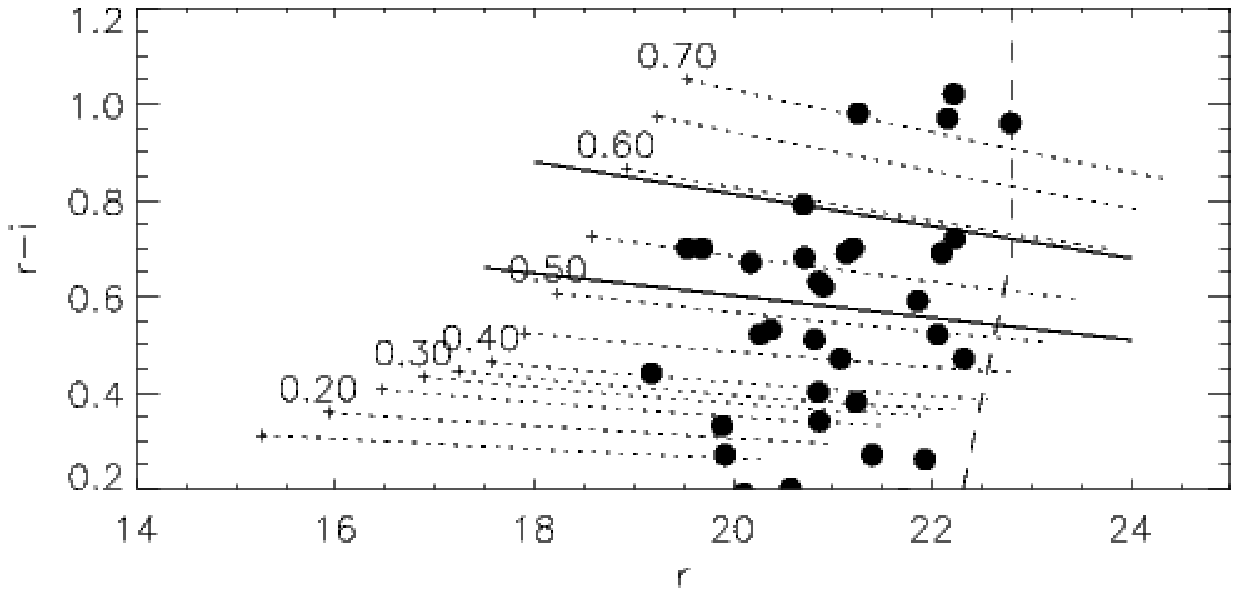,height=\cmheight\textheight}}}
\caption[Optical Image and CM Diagram of source \#161]{\footnotesize
{The top panel is the optical image of source \#161, centered on the X-ray
centroid; the bottom panel is the CM diagram for the region within the smaller overlaid circle at top.
The top image is an 1800~s Gunn
$r$ exposure with seeing $\sim1.8\arcsec$. The smaller circle indicates a 500~\hkpc{} radius centered
on the largest overdensity in the field,  at a redshift of $z=0.55$. We suggest that the more distant
structure dominates the X-ray emission (see text and Table \ref{tab_newclusters3}). The CM diagram for
the more distant cluster contains only those galaxies within the indicated 500~\hkpc~radius of the BCG,
for which we have plotted
$r-i$ color vs. $r$ magnitude. The dotted lines are galaxy color models based on the population
synthesis calculations of
\citet{kod97} and \citet{kod98}. Each line represents the range in color and magnitude for elliptical
galaxies of different mass observed in the appropriate filters at the labeled redshift.  Solid lines
show the estimated redshift range of the cluster red sequence, which includes a conservative photometric
error estimate of $\pm0.1$~mag in color. The dashed line indicates the limiting magnitude (5$\sigma$
detection limit) in this field.
\label{fig_161}
} 
} 
\end{figure*}

Source \#161:
In Fig. \ref{fig_161} we present the optical Gunn $r$ image of source \#161 (top panel), and the $r-i$
vs. $r$ CM diagram (bottom panel) for galaxies detected in the field of source \#161 within a radius of
500~\hkpc. This radius is shown in the top panel of Fig. \ref{fig_161} by the smaller solid circle,
centered on the probable BCG. This source exhibits a high overdensity of galaxies across the field;
however there appears to be more than one physical structure based on galaxy color and projected
density. The second structure is indicated in Fig. \ref{fig_161} by the larger solid circle (also of
radius 500~\hkpc~at the lower redshift of this structure; see below). Accordingly, our over-density
estimates have been corrected (to lower values) in an attempt to avoid contributions from galaxies at
different redshifts. We have segregated galaxies by color to assign them to each of the two
concentrations, and corrected our over-density estimates by the fraction of galaxies in each structure. 
We estimate that the most dominant CRS lies within the redshift range
$z=0.52-0.59$. In order to segregate possibly contaminating galaxies, we set a color limit of $r-i \geq
0.5$ for this structure.  Galaxies bluer than this limit were excluded from the counts used  for the
calculation of B$_{gc}$. At a redshift of
$z=0.55$, the new number of net galaxies lowers the measured B$_{gc}$ value from 2790~\bggmph, to
$1340\pm560$~\bggmph, which indicates a cluster consistent with ARC 1--2. The estimated X-ray flux from
a cluster with this overdensity (see Table \ref{tab_newclusters3}) is higher than the actual IPC
aperture fluxes observed so that this cluster is easily rich enough to produce the observed X-ray
emission.  The next richest CRS in the field lies in the redshift range
$z=0.32-0.38$. At a  redshift of
$z=0.35$, we measure a corrected galaxy over-density of B$_{gc}=740\pm240$~\bggmph, corresponding to
38\% of the flux expected from the more distant cluster. Thus, the more distant structure dominates the
expected X-ray flux. However, the apparent high redshift cluster is also
$90$~arcsec from our X-ray centroid, which suggests additional sources or extended emission could
contribute to the X-ray detection (see Paper 1). This cluster is somewhat irregular in appearance, with
only mild central concentration around the probable BCG and only a modest CRS. Due to the difference in
position of the apparent cluster and the X-ray centroid, the possibility of contamination by a second
cluster or an AGN, and the complex nature of our galaxy segregation and over-density calculations, we
identify source
\#161 as only a possible cluster of galaxies at $z=0.52-0.59$. We emphasize that this source presents a
complex situation, and our conservative possible identification reduces the impact this cluster has on
the EMSS sample (Paper 1).
\epsscale{1.0}

\begin{figure*}[ht]
\parbox{0.49\textwidth}{
\centerline{\psfig{figure=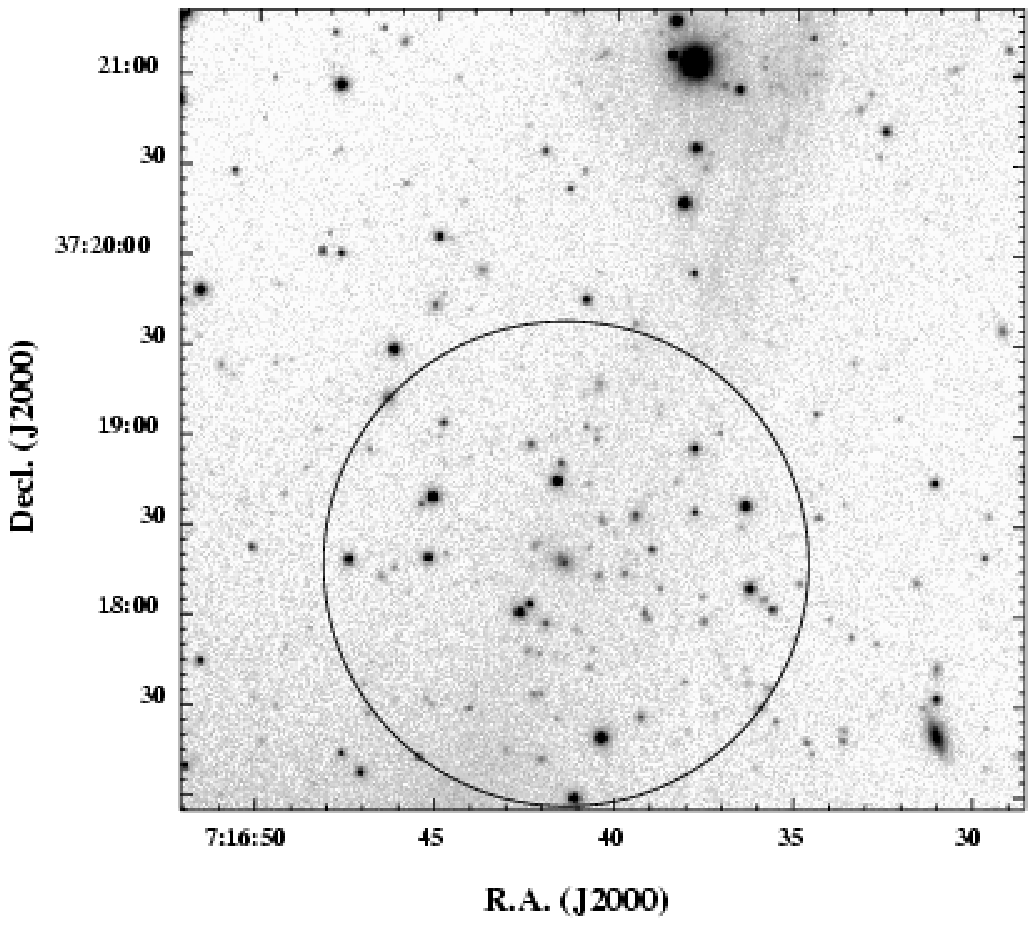,height=\rheight\textheight}}}
\parbox{0.49\textwidth}{
\centerline{\psfig{figure=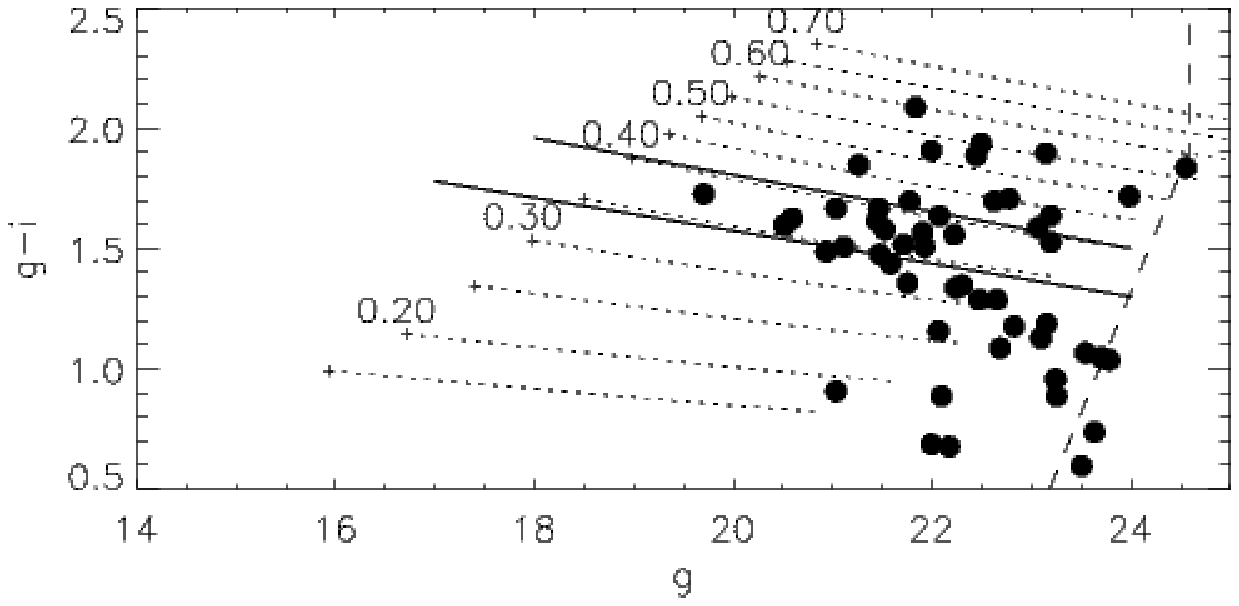,height=\cmheight\textheight}}}
\caption[Optical Image and CM Diagram of source \#1310]{The top panel is the optical image of source
\#1310 (a 3600~s Gunn $r$ exposure; seeing was
$\sim1.4\arcsec$), with a 500~\hkpc{} radius circle centered on the BCG. The bottom panel is the
CM diagram for galaxies within this radius. Annotations are identical to Fig. \ref{fig_161}.
\label{fig_1310}}
\end{figure*}

Source \#1310: In Fig. \ref{fig_1310} we present the Gunn $r$ image of source
\#1310 (top panel), and the $g-i$ vs. $g$ CM diagram (bottom panel) for galaxies detected in the field
of source \#1310 within a radius of 500~\hkpc, shown by a solid circle centered on the BCG. The CM
diagram for this field indicates a CRS with estimated redshift of
$z=0.34-0.40$.  At $z=0.37$, we measure B$_{gc}=1200\pm270$~\bggmph~(equivalent to ARC 1--2). 
The X-ray flux measured with the IPC is consistent with a cluster of this richness and redshift.
This cluster
has a moderate central concentration, with a clear BCG near the field center.
\epsscale{1.0}

\begin{figure*}[ht]
\parbox{0.49\textwidth}{
\centerline{\psfig{figure=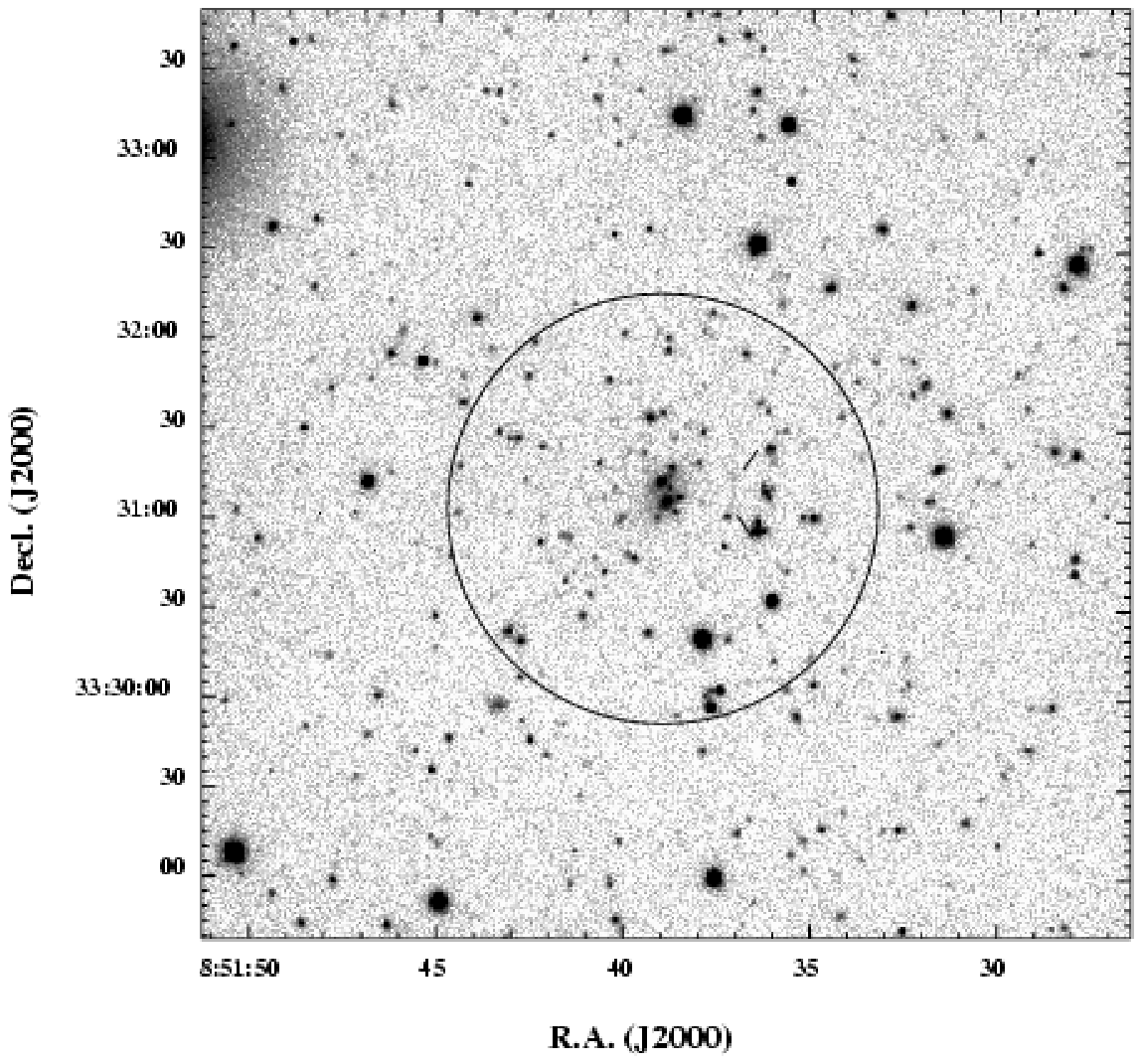,height=\rheight\textheight}}}
\parbox{0.49\textwidth}{
\centerline{\psfig{figure=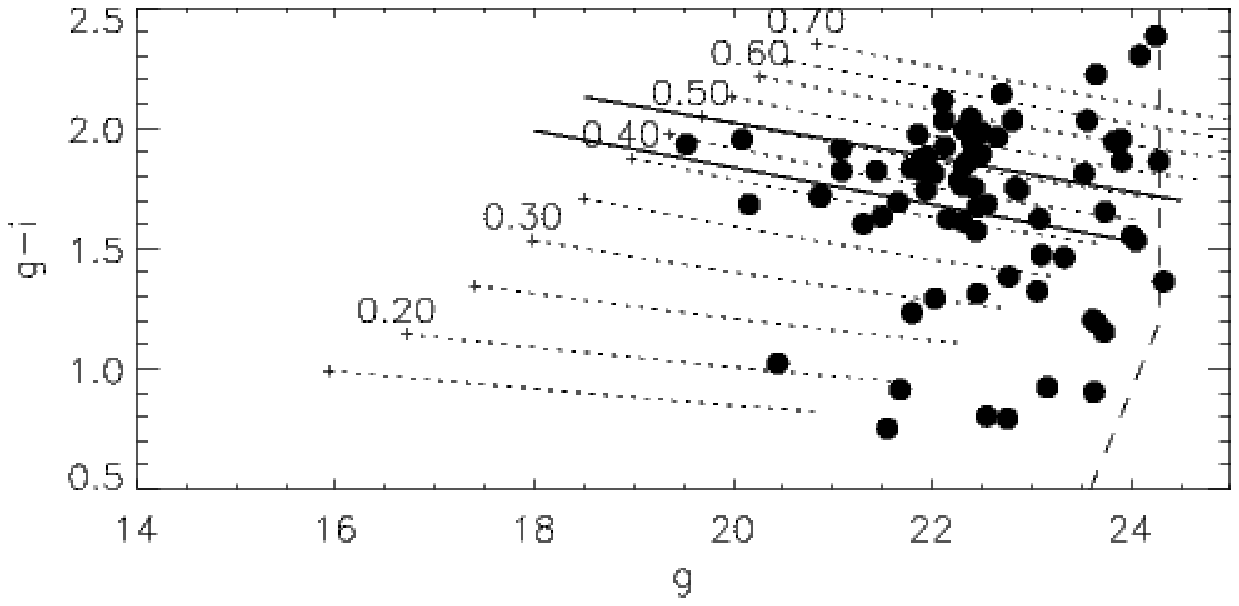,height=\cmheight\textheight}}}
\caption[Optical Image and CM Diagram of source \#1492]{The top panel is the optical image of source
\#1492 (an 1800~s Gunn $r$ exposure; seeing was
$\sim1.3\arcsec$), with a 500~\hkpc{} radius circle centered on the southern BCG. Note the faint arc
$\sim25\arcsec$ west of the BCGs. Our 3 color data indicates it is significantly bluer than the cluster
galaxies. The bottom panel is the CM diagram for galaxies within the indicated 500~\hkpc{} radius. 
Annotations are identical to Fig. \ref{fig_161}.
\label{fig_1492}
}
\end{figure*}

Source \#1492: In Fig. \ref{fig_1492} we present the Gunn $r$ image of source
\#1492 (top panel), and the $g-i$ vs. $g$ CM diagram (bottom panel) for galaxies detected in the field
of source \#1492  within a radius of 500~\hkpc,
shown by a solid circle centered on the southern galaxy in the pair of BCGs. This cluster has a
dense core and large overdensity of galaxies. Note the faint blue arc to the west of the BCGs, whose
$g-r$ color is $\sim2$ mags bluer than the BCGs. The blue arc is clear evidence for a deep gravitational
potential well, consistent with this source being a rich cluster of galaxies. From the CRS, we estimate
a redshift of
$z=0.42-0.50$ for this cluster.  At $z=0.47$, we measure
B$_{gc}=2240$ \bggmph~(equivalent to ARC 4), the highest galaxy richness in our sample.
\epsscale{1.0}

\begin{figure*}[ht]
\parbox{0.49\textwidth}{
\centerline{\psfig{figure=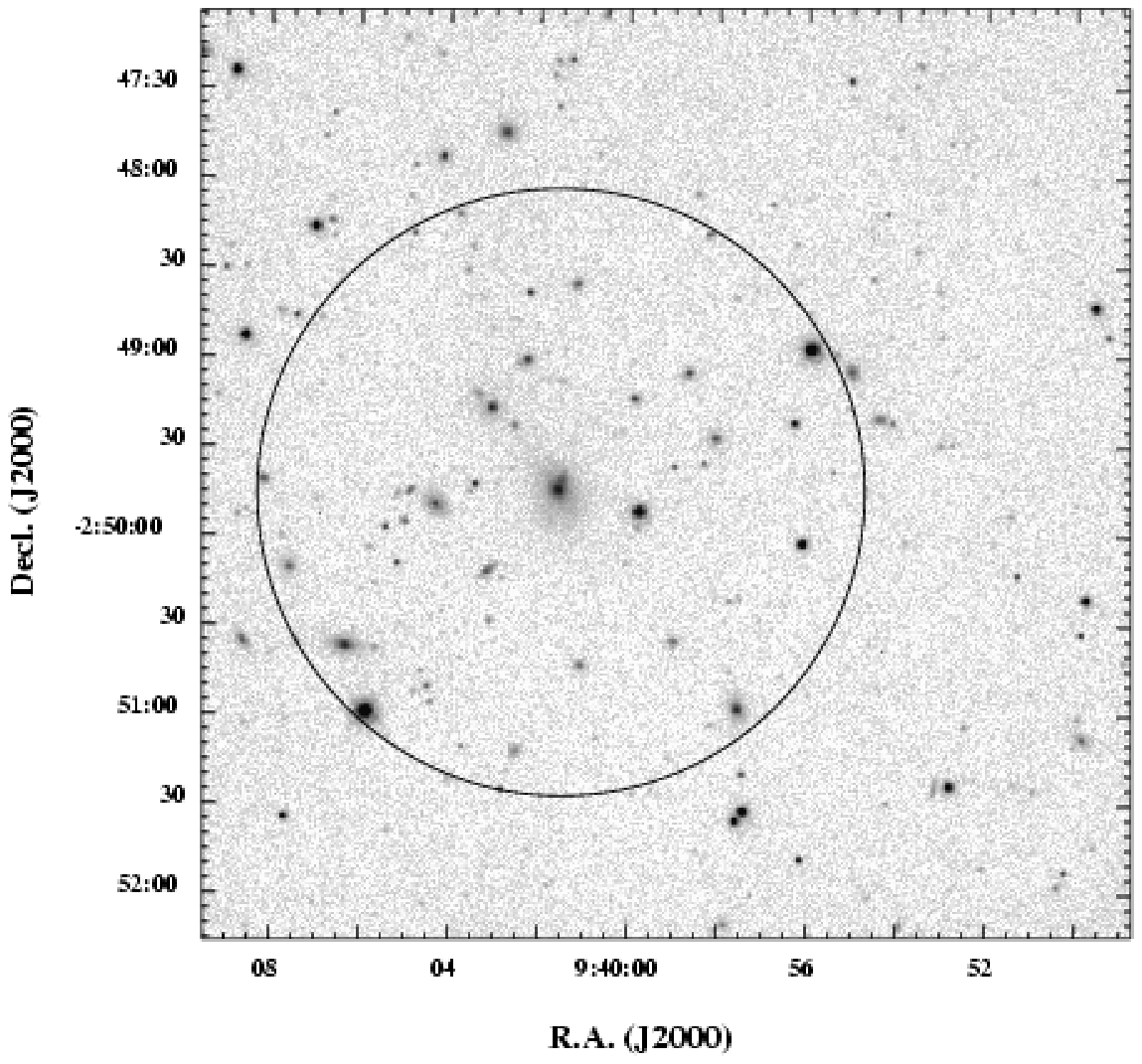,height=\rheight\textheight}}}
\parbox{0.49\textwidth}{
\centerline{\psfig{figure=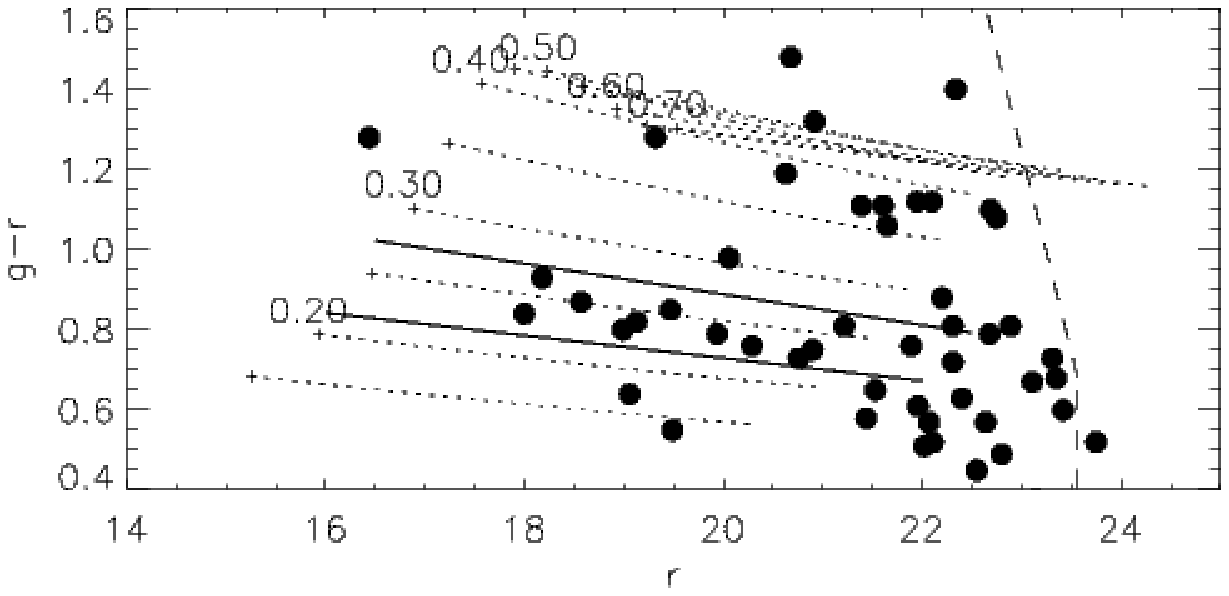,height=\cmheight\textheight}}}
\caption[Optical Image and CM Diagram of source \#1605]{The top panel is the optical image of source
\#1605 (an 1800~s Gunn
$r$ exposure; seeing was $\sim1.2\arcsec$), with a 500~\hkpc{} radius circle centered on the BCG.
Note the star projected onto the BCG. The bottom panel is the CM diagram for galaxies within the
indicated 500~\hkpc{} radius. Note that the bright galaxy which appears far redder than the red
sequence is in fact the BCG, for which the projected star has affected the galaxy
photometry. Annotations are identical to Fig.
\ref{fig_161}.
\label{fig_1605}
}
\end{figure*}

Source \#1605: In Fig. \ref{fig_1605} we present the Gunn $r$ image of source \#1605 (top panel),
and the $g-r$ vs. $r$ CM diagram (bottom panel) for galaxies detected in the field of source \#1605 
within a radius of 500~\hkpc, shown by a solid circle centered on the BCG. 
This cluster has an estimated redshift of
$z=0.22-0.27$.
At a redshift of $z=0.25$, we measure B$_{gc}=830\pm$ \bggmph~(ARC 0--1).
The inferred $L_X$ from the IPC flux is significantly lower than expected from a cluster of this
richness.
The galaxies in this cluster are distributed widely
(across $>5\arcmin$ in diameter), and it has a dominant cD-type galaxy with an extended stellar
envelope. The BCG has a star projected onto it which affects both its magnitude and color in the CM
diagram.
\epsscale{1.0}

\begin{figure*}[ht]
\parbox{0.49\textwidth}{
\centerline{\psfig{figure=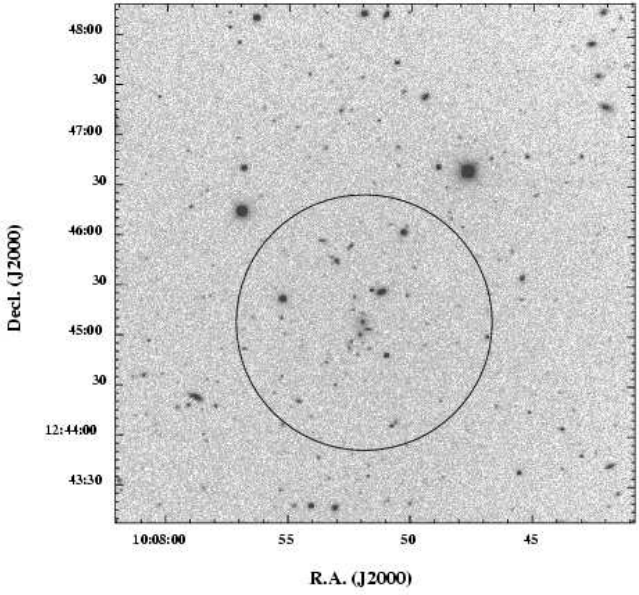,height=\rheight\textheight}}}
\parbox{0.49\textwidth}{
\centerline{\psfig{figure=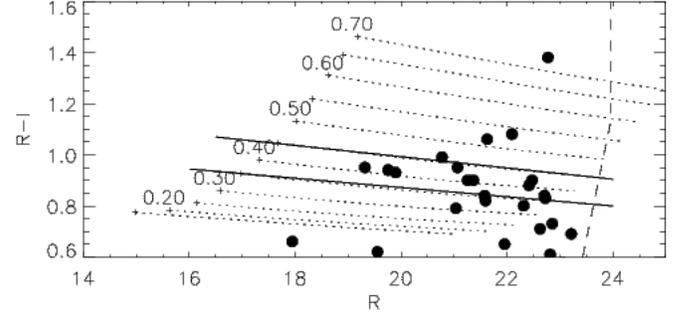,height=\cmheight\textheight}}}
\caption[Optical Image and CM Diagram of source \#1681]{The top panel is the optical image of source
\#1681 (an 1800~s Kron-Cousins $R$ exposure; seeing was $\sim1.2\arcsec$), with a 500~\hkpc{} radius
circle centered on the BCG. The bottom panel is the
CM diagram for galaxies within this radius. Annotations
are identical to Fig.
\ref{fig_161}.
\label{fig_1681}}
\end{figure*}

Source \#1681: In Fig. \ref{fig_1681} we present the Kron-Cousins $R$ image of source
\#1681 (top panel),
and the $R-I$ vs. $R$ CM diagram for
galaxies detected in the field of source \#1681 within a radius of 500~\hkpc,
shown in the top panel by a solid circle centered on the BCG. This cluster has an estimated redshift of
$z=0.35-0.45$.
At a redshift of $z=0.41$, we measure B$_{gc} = 600$ \bggmph (ARC $0$).
The flux from the IPC for this source is
consistent with this cluster being the sole source of the X-rays in the field. The
optical appearance is of a compact core which is confirmed by the CM diagram. There are also a large
number of galaxies falling along the red sequence but distributed much farther from the core, in a
shallow extended envelope.
\epsscale{1.0}

\begin{figure*}[ht]
\parbox{0.49\textwidth}{
\centerline{\psfig{figure=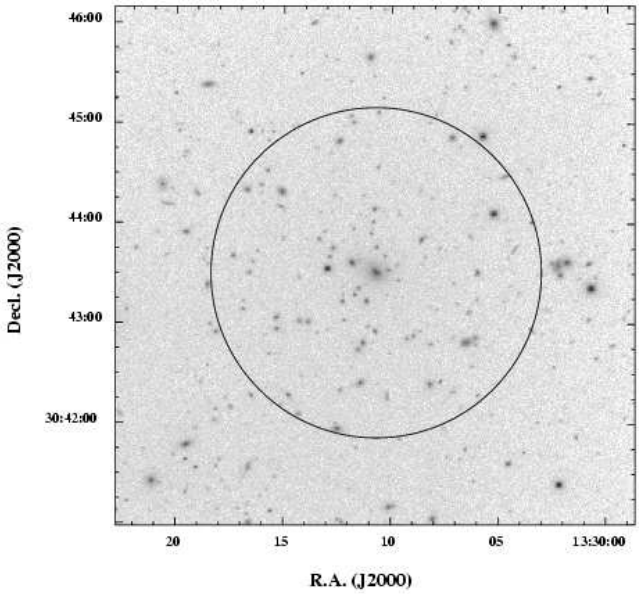,height=\rheight\textheight}}}
\parbox{0.49\textwidth}{
\centerline{\psfig{figure=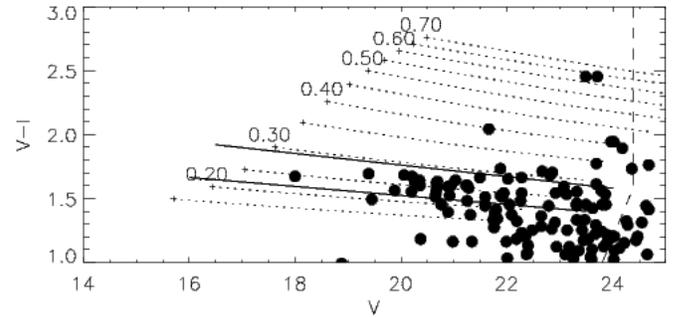,height=\cmheight\textheight}}}
\caption[Optical Image and CM Diagram of source \#2436]{The top panel is the optical image of source
\#2436 (a 2400~s Kron-Cousins $R$ exposure with
seeing was $\sim0.9\arcsec$), with a 500~\hkpc{} radius
circle centered on the BCG. The bottom panel is the
CM diagram for galaxies within this radius. 
Annotations are identical to Fig.
\ref{fig_161}. Note that there was a second cluster detected in the field, but it is outside of the area
shown in the top panel. It is also not within the region of the CM diagram for the primary CRS shown at
bottom.
\label{fig_2436}}
\end{figure*}

Source \# 2436: In Fig. \ref{fig_2436} we present the Kron-Cousins $R$ image of source
\#2436 (top panel),
and the $V-I$ vs. $V$ CM diagram  (bottom panel) for
galaxies detected in the field of source \#2436 within a radius of 500~\hkpc,
shown in the top panel by a solid circle centered on the BCG. This source is coincident
with a \citet*{gun86} cluster with an unknown redshift. Our imaging easily detects this cluster and the
CM diagram indicates a CRS of redshift $z=0.22-0.28$.
At $z=0.26$, we measure B$_{gc} = 1450$ \bggmph (ARC 2).
There is a second, weaker, CRS present in the CM diagram, centered
$\sim 5\arcmin$ NE of the X-ray centroid, at a similar, but slightly higher redshift of
$z\sim0.3$, and an overdensity of B$_{gc} = 400-650$ \bggmph.
At this location and richness, it can only
contribute to the 3rd or 4th IPC apertures, and even then is likely to be only a small fraction of the
flux compared to the primary cluster. For this source, the smallest IPC aperture data alone is a
4.5$\sigma$ detection (see Paper 1), which is due solely to the $z=0.26$ cluster, and we identify this
source as a rich cluster of galaxies. Based on the galaxy colors of the two structures, which differ by
0.4 mag in $V-R$, it is unlikely that they are physically related.
\epsscale{1.0}

\begin{figure*}[ht]
\parbox{0.49\textwidth}{
\centerline{\psfig{figure=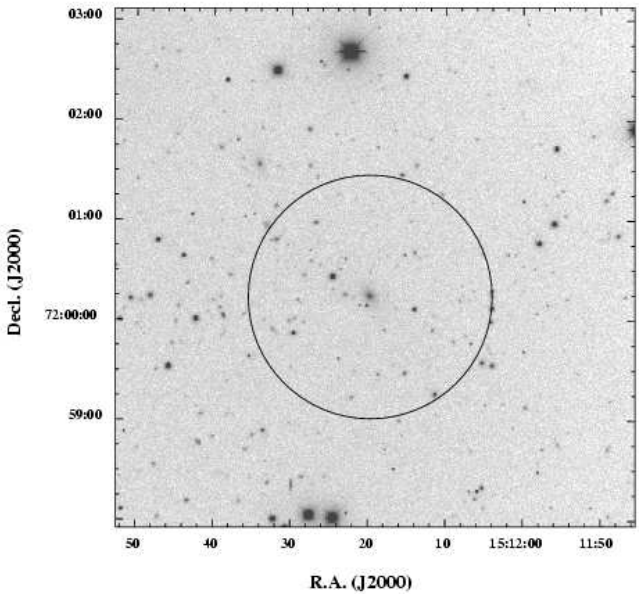,height=\rheight\textheight}}}
\parbox{0.49\textwidth}{
\centerline{\psfig{figure=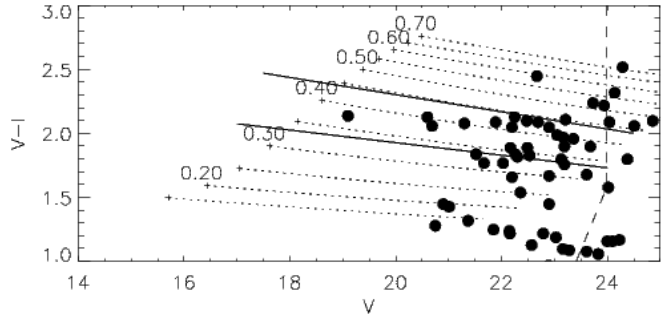,height=\cmheight\textheight}}}
\caption[Optical Image and CM Diagram of source \#2844]{The top panel is the optical image of source
\#2844 (an 1800~s Kron-Cousins $R$ exposure; seeing was $\sim1.0\arcsec$), with a 500~\hkpc{} radius
circle centered on the BCG. The bottom panel is the
CM diagram for galaxies within this radius. Annotations are identical to Fig.
\ref{fig_161}.
\label{fig_2844}}
\end{figure*}

Source \#2844: In Fig. \ref{fig_2844} we present the Kron-Cousins $R$ image of source
\#2844 (top panel),
and the $V-I$ vs. $V$ CM diagram (bottom panel) for
galaxies detected in the field of source \#2844 within a radius of 500~\hkpc,
shown in the top panel by a solid circle centered on the BCG. This field may contain a
cluster of galaxies, but our photometry for it is not self-consistent: the $V-R$ and $R-I$ data give
discrepant estimates of the color of the apparent CRS at the $0.05-0.1$~mag level. We experienced a
wide variation in seeing during these observations, which proved to be problematic for the subsequent
color measurements. Our best estimate (using the optimum constraint provided by the
$V-I$ data) is a CRS at redshift
$z=0.33-0.44$.
At $z=0.38$, we measure B$_{gc}=820$ \bggmph (ARC 0--1).
The expected X-ray flux is somewhat larger than the IPC aperture fluxes,
indicating this cluster is easily responsible for our X-ray detection. The
optical image reveals a dominant central galaxy, but otherwise very little central concentration. The CM
diagram indicates a CRS, though the galaxies within it are somewhat widely spatially distributed.
Due to the inconsistencies in our photometry, we identify this field as only a possible cluster of
galaxies.
\epsscale{1.0}

\begin{figure*}[ht]
\parbox{0.49\textwidth}{
\centerline{\psfig{figure=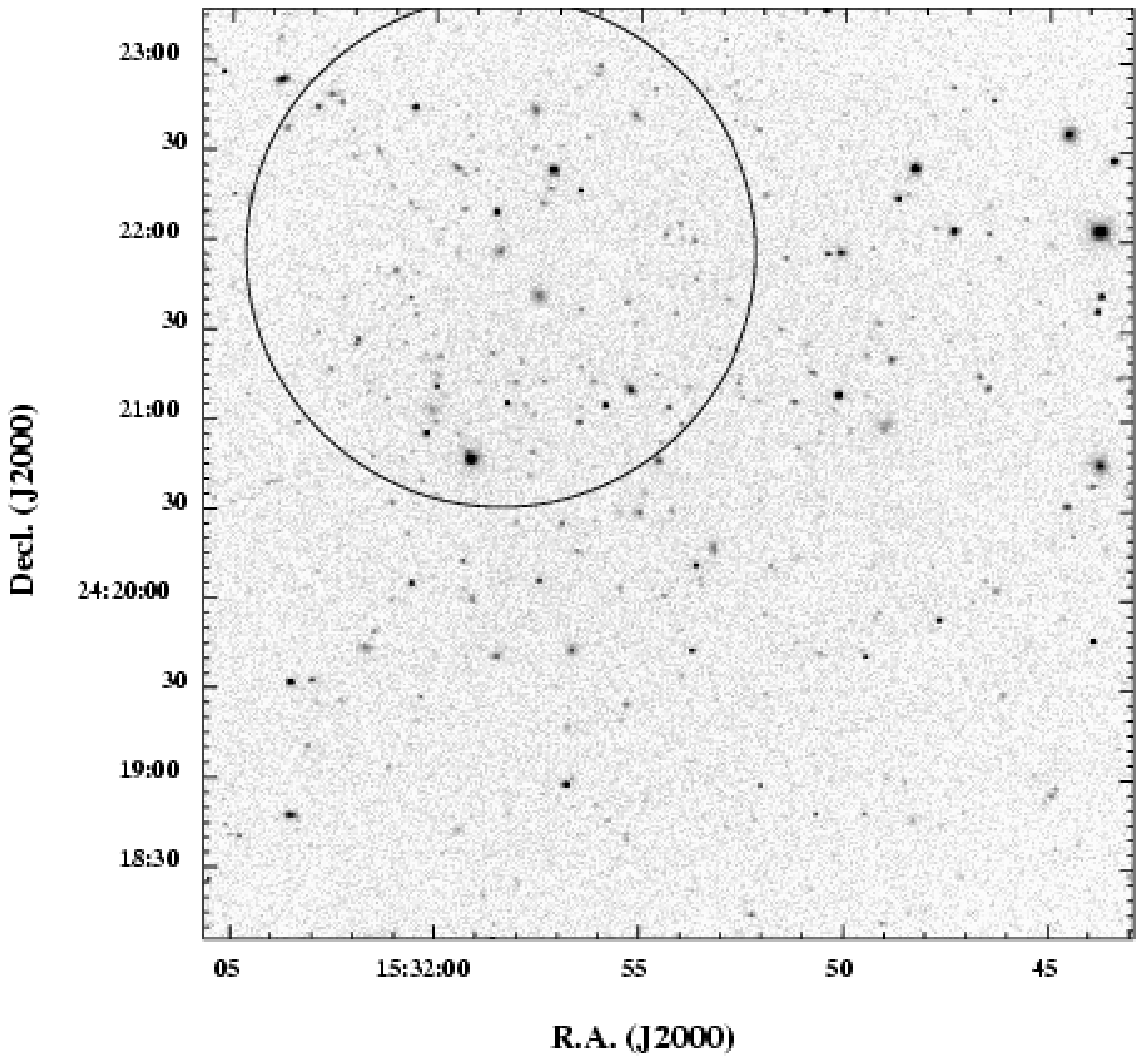,height=\rheight\textheight}}}
\parbox{0.49\textwidth}{
\centerline{\psfig{figure=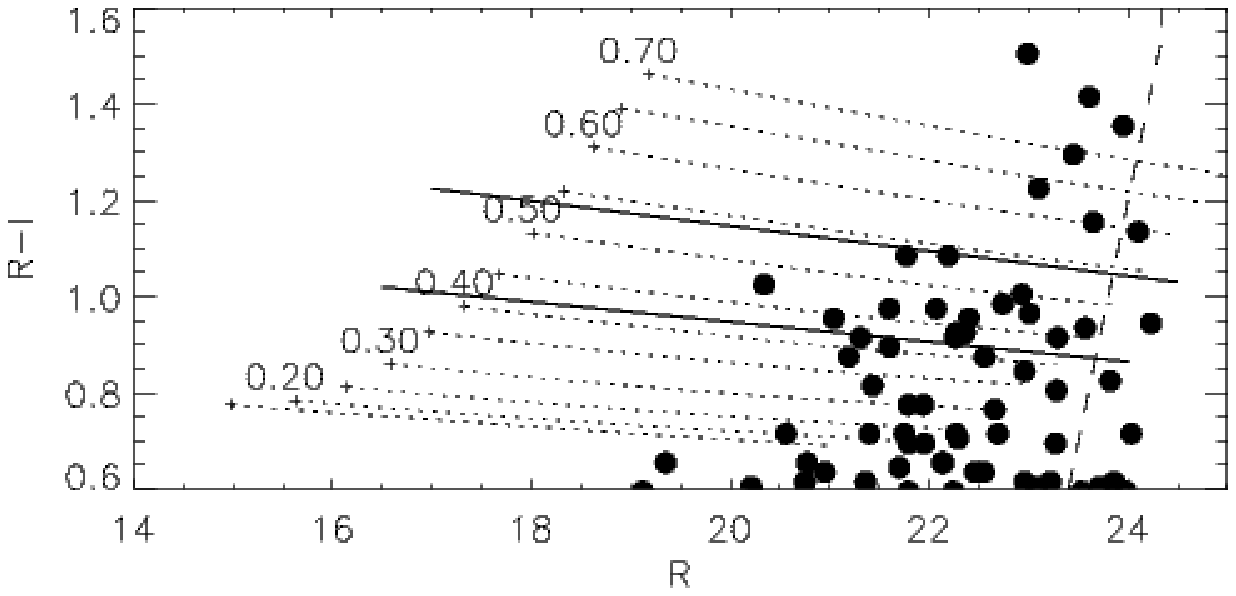,height=\cmheight\textheight}}}
\caption[Optical Image and CM Diagram of source \#2906]{The top panel is the optical image of source
\#2906 (an 1800~s Kron-Cousins $R$ exposure; seeing was $\sim0.8\arcsec$), with a 500~\hkpc{} radius
circle centered on the BCG. Note the bright foreground
galaxy 20\arcsec{} SW of the BCG at the center of the solid circle - we consider it possible that the
galaxy overdensity in this field is enhanced due to the contamination of galaxies at different
redshifts. Alternatively, this may be a cluster in the process of formation. The bottom panel is the CM
diagram for galaxies within 500~\hkpc{} of the BCG. Annotations are identical to Fig.
\ref{fig_161}.
\label{fig_2906}}
\end{figure*}

Source \#2906: In Fig. \ref{fig_2906} we present the Kron-Cousins $R$ image of source
\#2906 (top panel), and the $R-I$ vs. $R$ CM diagram (bottom panel) for galaxies detected in the field
of source \#2906 within a radius of 500~\hkpc, shown in the top panel by a solid circle centered on the
BCG. This field presents a significant galaxy overdensity, and an apparent CRS in the CM, but the BCG
is 2\arcmin~NE of the X-ray centroid, and there is a lack of concentration of galaxies around it.
Additional scrutiny of the galaxy colors implies possible contamination from galaxies at a variety of
redshifts, though there is no second CRS present (as in the case of \#161). We estimate the primary CRS
to be at a redshift of
$z=0.42-0.53$. At $z=0.47$, we measure B$_{gc} = 970$ \bggmph (ARC 1). Due to the offset of the BCG
from the X-ray centroid, and the possibility of contamination, we identify this source as only a
possible cluster of galaxies. In addition, this cluster has the highest blue fraction of galaxies
(defined below) in our sample, indicating either that this source is optically contaminated (and thus
not as rich as the B$_{gc}$ value indicates) or that it contains many late-type cluster galaxies, due
to an unevolved dynamical state, indicative of a  cluster still in formation. 
\epsscale{1.0}

Source \#3353: In Fig. \ref{fig_3353r} we present the Kron-Cousins $R$ image of source
\#3353. This field contains an obvious low-redshift ($z<0.1$) group or poor cluster of galaxies
extending well across our $10\arcmin \times 10\arcmin$ field. Because our redshift estimates are not
at all accurate below $z\sim0.15$ we have not tried to measure the redshift or galaxy overdensity
of this group. The measured IPC flux in the largest aperture indicates a  maximum luminosity of L$_X
\leq 1.0\times10^{44}$ ergs s$^{-1}$ for $z \leq 0.1$; given the magnitude of these galaxies a redshift
much less than 0.1 is suggested, so that an $L_X$ typical of an elliptical-rich galaxy group is implied
\citep{mul98}. There is a very large, dominant elliptical galaxy near the field center, with many
smaller elliptical and S0 galaxies evident across the entire field. It is very likely to be the source
of extended X-rays in this field, and therefore we identify it as a nearby group or poor cluster.
\begin{\myfigure}
\epsscale{\rscale}
\plotone{\figdirthree 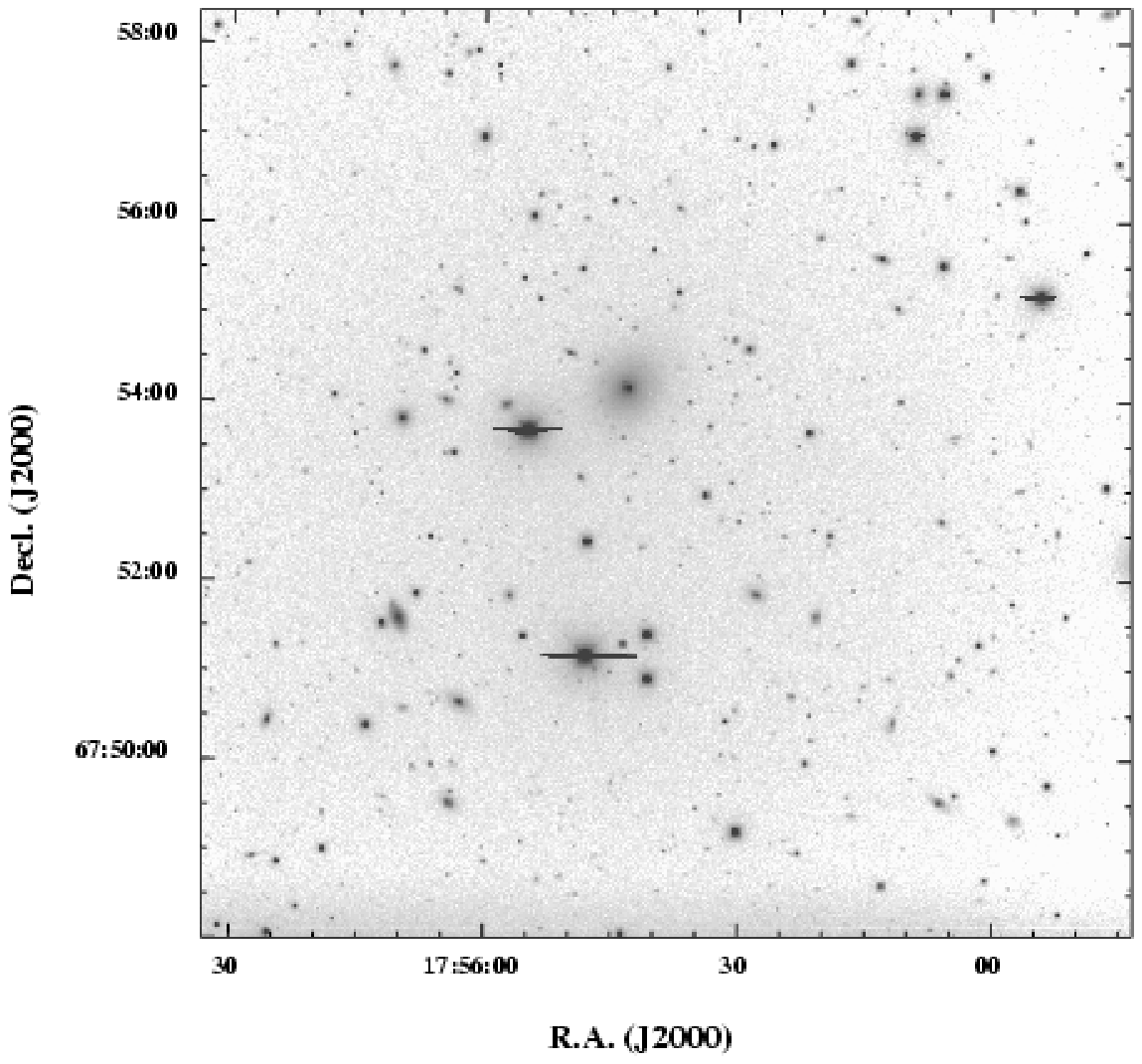} 
\caption[Optical Image of source \#3353]{Optical image of source \#3353 (an 1800~s
Kron-Cousins $R$ exposure;
seeing was $\sim1.2\arcsec$).
\label{fig_3353r}}
\end{\myfigure}
\epsscale{1.0}

\section{Discussion\label{sec_discussionp3}}

We have found several bona-fide clusters of galaxies within
the EMSS sky area that have gone previously undetected.
As shown in Paper 1, these clusters have significantly
extended X-ray emission (see Figure 3, Paper 1),
but are they truly low surface brightness (LSB) clusters? 
The modest spatial resolution of the IPC instrument, and the low S/N of these detections does not allow
us to make precise surface brightness profiles to answer this question conclusively, but we analyze the
data from the four X-ray apertures in an attempt to address the issue.
It is possible that these clusters are similar to other EMSS clusters, and were only missed by the
original EMSS because that sample selected against asymmetric objects (e.g., \citealt{ebe00} discuss
\#420, one of the clusters also rediscovered in this work; see Paper 1) near the flux limit of any given
observation, where most of these clusters are detected.  Alternatively, they may represent a new class
of clusters which lack any central X-ray concentration although they are massive. If these clusters are
in a different dynamical state from relaxed systems, they may exhibit the following observational
properties: low X-ray luminosity relative to their mass or richness, or a large fraction of blue,
presumably spiral, star-forming galaxies.
Either observation could be interpreted as evidence for clusters in the process of
formation, which could significantly affect cosmological studies of galaxy clusters. 
In this section we will scrutinize the available optical
and X-ray data for the sample as a preliminary investigation of their global properties.

Although the available IPC data is of low S/N, we can use our multiple aperture detections to
make a simple surface brightness analysis. We take the ratio of X-ray flux in the third to the first
IPC apertures (8.4 and 2.5 arcmin in diameter, see Paper 1), for the new clusters and compare it with
the same measurements made for the EMSS/CNOC clusters \citep[Canadian Network for Observational
Cosmology,][]{yee96} detected by our algorithm.
The EMSS/CNOC clusters are a subset of the EMSS cluster sample, which are
primarily rich, relaxed, clusters with high X-ray luminosity \citep[excepting
MS~$1621.5+2640$ and MS~$0906.5+1110$, which have complex X-ray and galaxy velocity
structures;][]{mor98,L99}.  In Figure
\ref{fig_apcompare} we present the ratio of third to first IPC aperture flux vs. redshift for the new,
optically discovered EMSS clusters (filled circles) and the EMSS/CNOC clusters\footnote{ We note that 3
of the 16 EMSS/CNOC clusters were not detected by our IPC reanalysis because they lay outside the
central region of the IPC detector (see Paper 1). Two of these clusters were detected by the
IPC as extended objects, and we have used the ratio of the EMSS ``Extended Counts'' (see
\citealt{gio90a}, Table 3, Column 9) to the EMSS detect cell counts as an approximation to the
aperture ratios (open triangles in Figure \ref{fig_apcompare}). Comparing the two methods for the
other CNOC clusters seen as extended by the IPC shows consistency with our aperture ratios
to within
$\sim20\%$.}  (open circles and open triangles), while filled triangles represent additional new
clusters in the EMSS found outside our optical imaging campaign (e.g., clusters found by other groups;
see Paper 1). The dotted curve represents the expected flux ratio for a cluster whose X-ray emission is
characterized by a standard $\beta$-model with $\beta = 2/3$ and
$r_{core} = 250$\hkpc. We can see that many of the new clusters (filled symbols) have a greater flux
ratio than the CNOC clusters. A Kolmogorov-Smirnov test indicates that the probability the two sets of
flux ratios are drawn from the same population is 0.8\%. Several of the new clusters lie well above
the dotted line, indicating their X-ray emission has less central concentration than expected \citep[see
e.g.,][]{hen92_h92}. This is our strongest indication that some of the new clusters have a different
surface brightness distribution than the other X-ray selected clusters in the EMSS sample, and this is
why they were missed by the EMSS detection algorithm. Although a definitive measure of the surface
brightness distributions requires higher resolution and S/N X-ray imaging, from Figure
\ref{fig_apcompare} we can see that many of these clusters represent an observationally distinct
class of objects.
\begin{\myfigure}
\epsscale{\figscale}
\plotone{\figdirthree 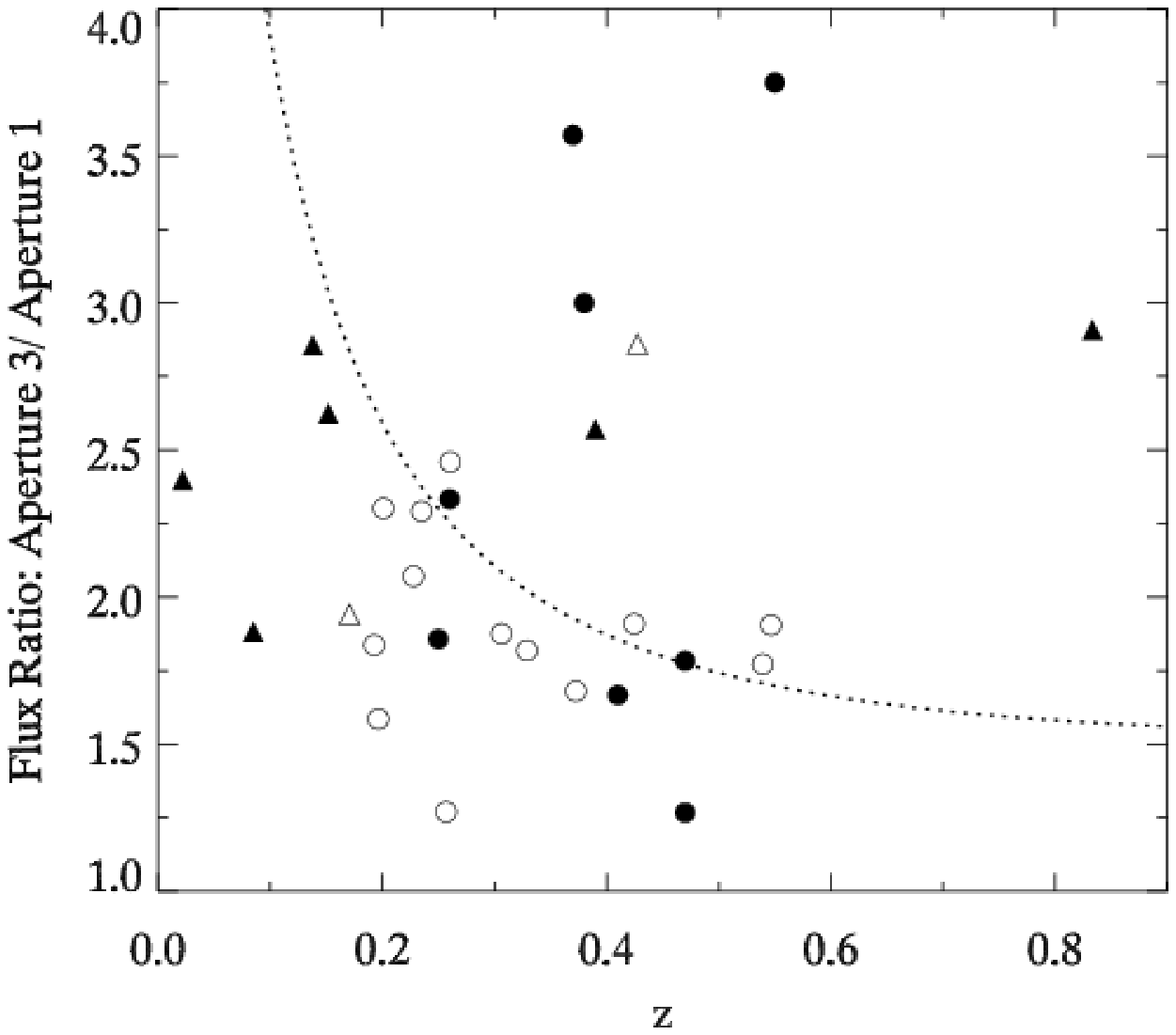}
\caption[Ratio of IPC Aperture Three to Aperture One Flux vs. Redshift]{Ratio of IPC aperture three to
aperture one flux vs. redshift for the new, optically discovered EMSS clusters described in
this paper (filled circles) and clusters in the EMSS/CNOC sample (open circles). Additional new
clusters discovered by other groups (see Paper 1) are shown as filled triangles. The
dotted line represents the expected flux ratio for clusters whose emission can be described by
$\beta=2/3$ and $r_{core} = 250$ \hkpc. The values for two CNOC clusters (MS~$0906.5+1110$,
MS~$1621.5+2640$), were calculated differently (see text); they are shown as open triangles. Errors on
the flux measurements are typically
$15-25\%$.}
\label{fig_apcompare}
\end{\myfigure}
\epsscale{1.0}

There are two possibilities which would result in the
high IPC aperture flux ratios we report for these new clusters. We
can illustrate these possibilities by looking at two individual
clusters: (1) MS~$1621.5+2640$, 
an original EMSS cluster which is part of the
CNOC sample, and (2) Cl~J0152.7-1357, discovered both in this
survey and in the Wide Angle \rosat{} Pointed Survey (WARPS; this
cluster is \#420 in our new IPC catalog). In the X-rays,
MS~$1621.5+2640$ has an extremely diffuse X-ray structure, no significant central peak, and only a
small fraction of the expected IPC flux is detected in
a long (43 ksec) \rosat{} HRI exposure, the remainder presumably not detectable above the high HRI
background
\citep{L99}. This cluster exhibits a large aperture flux ratio (2.9, see Figure \ref{fig_apcompare}, it
lies at
$z=0.4274$), a relatively large blue galaxy fraction, and significant optical evidence
of a recent merger
\citep{mor98}. On the other hand, in a \rosat{} PSPC exposure
Cl~J0152.7-1357 shows a significant central concentration, but
with a double-peaked structure and some apparent asymmetry (see
the detailed X-ray analysis of \citealt{ebe00}). Examining Fig.
\ref{fig_apcompare}, we see that Cl~J0152.7-1357 ($z=0.83$), has a flux ratio exceeding the highest
value for EMSS/CNOC clusters, including MS~$1621.5+2640$.  And yet
the \citeauthor{ebe00} analysis shows that it is not truly LSB in
X-rays; instead, they characterize it as moderately asymmetric, though significantly unrelaxed, and
apparently merging. We also note that
$T_X=6.5^{+1.74}_{-1.19}$~keV for Cl~J0152.7-1357 \citep{del00},
while $T_X$ for the cluster MS~$1054.4-0321$ is significantly higher ($10.4^{+1.7}_{-1.5}$~keV;
\citealt{jel01}), despite these two clusters having nearly identical $L_X \sim 1.5 \times 10^{45}$ \lx{}
in the
\eband{} band, and $z=0.83$; i.e., Cl~J0152.7-1357 may be less virialized than
MS~$1054.3-0321$, but is nonetheless an extremely luminous addition to the EMSS cluster sample. We
expect that the uninvestigated new clusters will either be truly LSB, similar to MS~$1621.5+2640$, or
moderately asymmetric, similar to Cl~J0152.7-1357. 

So, while these aperture flux ratios are not
definitive indicators of LSB clusters, they do seem to indicate some differences in X-ray morphology.
The real question remains, though, are the new clusters physically distinct in structure and galaxy
populations as well?
We have previously shown (Figures 4 \& 5, Paper 1) that the redshifts and X-ray luminosities of
the new clusters are distributed similarly to the original distant ($z>0.14$) EMSS cluster sample of
\citet{hen92_h92}. In Figure
\ref{fig_bhistcompare} we present the galaxy over-density
distribution for all new clusters in our catalog with measured B$_{gc}$ values, compared to the B$_{gc}$
values for clusters in the EMSS/CNOC sample \citep{yee01}. We note that of the nine clusters presented
in this paper, only
\#3353 does not have an estimated redshift necessary to estimate B$_{gc}$ accurately, and is not shown.
We see that the newly-discovered clusters occupy a similar range of B$_{gc}$ values as the rich cluster
sample. 
\begin{\myfigure}
\epsscale{\figscale}
\plotone{\figdirthree 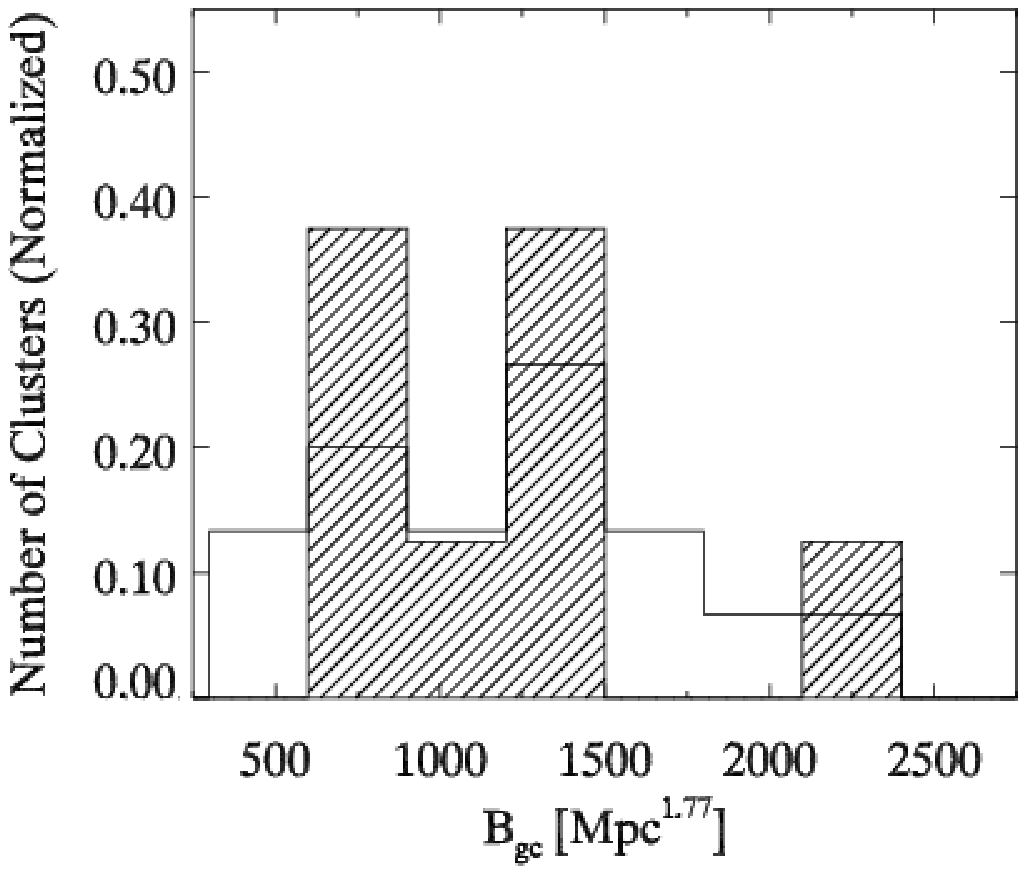}
\caption[Normalized B$_{gc}$ Distribution]{The normalized B$_{gc}$ distribution of both newly-discovered
EMSS clusters (shaded histogram) and clusters in the CNOC sample (open histogram; data from
\citealt{yee01}).}
\label{fig_bhistcompare}
\end{\myfigure}
\epsscale{1.0}

An empirical B$_{gc}-L_X$ relation has been created for the CNOC clusters \citep{yee01}, showing a
mild correlation between the two quantities. In Fig.
\ref{fig_bgc_lx} we present B$_{gc}$ vs. log
$L_X$ for the new clusters, with the CNOC B$_{gc}-L_X$ relation shown as a dotted line.
This figure shows that the new clusters exhibit a wide scatter, but generally conform to the relation,
and do not appear to represent a distinct population, though we caution that so few objects are not a
definitive investigation. Clusters \#1605 and \#2436 exhibit the most significant departure from
the B$_{gc}-L_X$ relation, with apparently low X-ray luminosity relative to their richness. However,
these two objects are not outliers in Figure \ref{fig_apcompare} or in the other comparisons we present
in this paper (see below).
\begin{\myfigure}
\epsscale{\figscale}
\plotone{\figdirthree 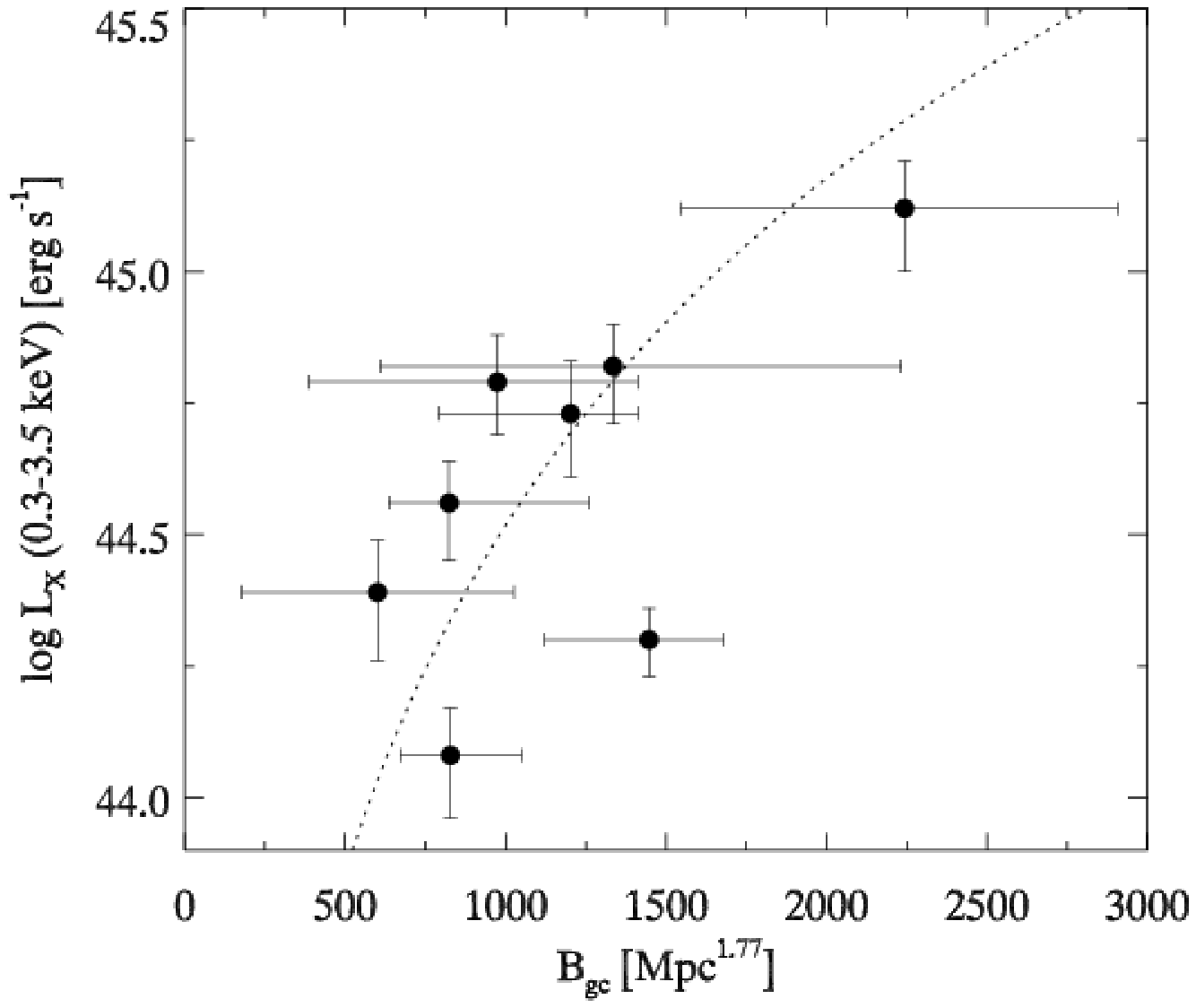}
\caption[B$_{gc}$ vs. $L_X$ for the New Clusters]{B$_{gc}$ vs. $L_X$ in the
\eband~band for the new clusters presented in this paper (excluding
\#3353, for which B$_{gc}$ data is unavailable). The dotted line represents the empirical B$_{gc} - L_X$
relation based on the CNOC cluster sample of
\citet{yee01}.}
\label{fig_bgc_lx}
\end{\myfigure}
\epsscale{1.0}

A parameter used as an indicator of evolution in cluster populations is the blue galaxy fraction
($f_b$). The original definition of $f_b$ was based on the fraction of galaxies in a cluster which
are more than 0.2 mag bluer in $B-V$ than the early-type galaxies \citep{but78}. The observation
that rich clusters exhibit higher values of $f_b$ with increasing redshift (the Butcher-Oemler
effect) is generally thought to be caused by a systematic gradient in the evolutionary state of
clusters. Two physical processes are likely responsible for the effect. 
Either the
mechanism which quenches star formation in cluster galaxies
is less effective at high
redshift, or the infall rate of blue galaxies into cluster potentials increases
with redshift
\citep[see e.g.;][]{dre83,aba99,bal99,ell01,kod01}. Both processes indicate that at high $z$ we are
seeing cluster populations undergoing continuing evolution. An investigation of $f_b$ in our new
clusters is therefore compelling: if we see higher than average
$f_b$ values in the new clusters presented here, this would indicate a class of clusters in an earlier
evolutionary state, although contemporaneous with the previously known systems.

Here we define the blue galaxy fraction similarly to the method of
\citet{ell01}, although with a few significant differences (see below). Based on the slope and zeropoint
of a line in the cluster CM diagram approximating the CRS, we make two color cuts and a magnitude cut to
count red and blue galaxies which may be cluster members. First, galaxies in each field outside of a
0.5~\hmpc{} radius are excluded.
Second, galaxies
fainter than either the $k$- and evolution-corrected absolute magnitude of $M_r^{k,E} = -19$, or the
completeness limit (taken to be 0.8 mags brighter than the
$5\sigma$ detection limit) for each filter are excluded. The $K$-corrections are calculated from
\citet{col80}, and the evolution is assumed to be $\Delta M_r^{k,E} = - z$ \citep{lin99}. Third, the
slope and zeropoint of a hypothetical, more distant CRS than the the cluster is defined as
the red limit for inclusion. Specifically, galaxies redder than a CRS at ($z_{cluster}+0.1$) are
considered likely background contamination and excluded. Finally, a blue limit is defined as a line in
the CM diagram having the slope of the cluster CRS, but with a zeropoint bluer by one-half the color
difference between an E galaxy and an Sbc galaxy in the filter and color combination observed, where
the galaxy colors are given by
\citet{fuk95}. These cuts effectively divide the CM diagram for each cluster into three components:
red background galaxies ($N_{red}$), red galaxies assumed to be associated with the CRS ($N_{CRS}$),
and blue galaxies which are an indistinguishable combination of foreground contamination and blue
cluster members ($N_{blue}$). In Fig. \ref{fig_blue_frac_cm_1605} we show a representative CM
diagram in $g-r$ vs. $r$ illustrating these color cuts for the galaxies in cluster \#1605.
\begin{\myfigure}
\epsscale{\figscale}
\plotone{\figdirthree 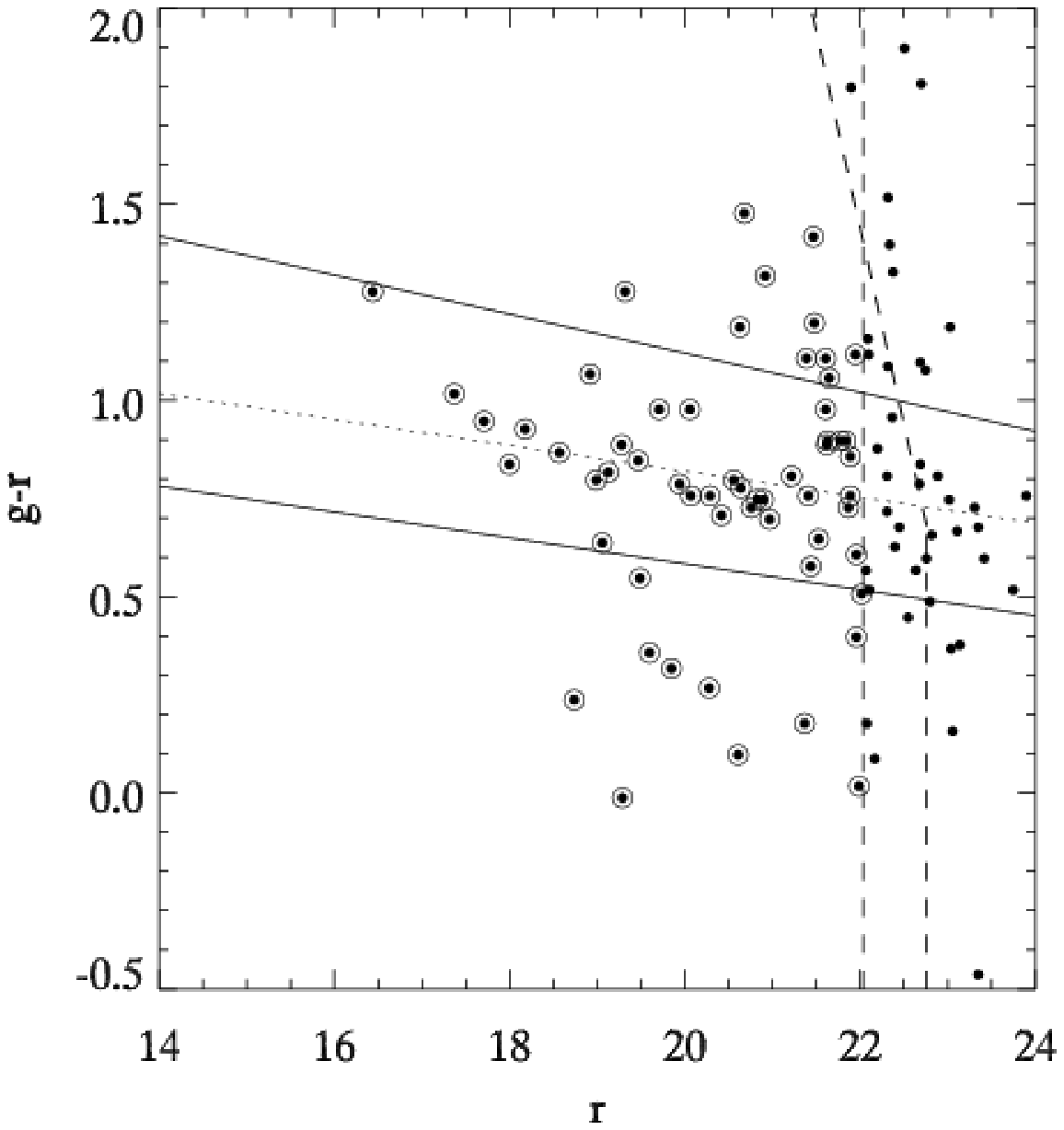}
\caption[CM diagram
for cluster \#1605, with the Red and Blue galaxy Color Cuts]{CM diagram
for galaxies in cluster \#1605, with the red and blue color cuts described in
the text overlaid as solid lines. The dotted line shows the predicted CRS from the
\citet{kod97} models at our estimated redshift of
$z=0.25$, while the broken dashed line indicates the completeness limits for this field and the vertical
dashed line indicates the $M_r^{k,E} = -19$ limit. Galaxies within 0.5~\hmpc{} radius of the BCG are
shown with dots; those which are above the brightest of the magnitude limits are circled.}
\label{fig_blue_frac_cm_1605}
\end{\myfigure}
\epsscale{1.0}

These photometrically defined color cuts result in galaxy counts that are contaminated by field
galaxies, and are thus a less robust evolutionary indicator than those derived from an
unbiased spectroscopically selected sample of confirmed cluster members (such as that defined for the
CNOC clusters by
\citealt{ell01}). Nonetheless, we can improve our counts in each component by statistically subtracting
the expected field galaxy contribution. We make a CM diagram in the same filter and color combination,
and magnitude limits as each cluster for the CNOC field galaxy observations of patch 0223+20
\citep{yee00}. These observations of 1,409 square arcmin are a precisely and accurately defined set
of over 18,000 field galaxies observed at CFHT in five colors, to a depth of $R\sim24.0$. The CNOC
field galaxy CM is then divided into precisely the same three components as the cluster CM diagram, and
the resulting galaxy counts are scaled to a sky area of 0.5~\hmpc{} radius as observed at each cluster's
redshift.

After subtracting these background counts from each component, 
we define 
\begin{equation}
f_b = \frac{N_{blue}}{(N_{CRS} + N_{blue})}.
\end{equation}
We note that the background subtracted red excess component, $N_{red}$, should ideally be zero; for
our sample of new clusters, $N_{red}$ has a mean value of only
$1.4\pm1.4$ galaxies, indicating that we have proper photometric calibration between our observations
and the CNOC data, and therefore an accurate field subtraction.  We also calculated this
photometric
$f_b$ 
for 15 CNOC clusters to compare the two
samples, and the mean $N_{red}$ value for CNOC clusters was $1.9\pm0.7$ galaxies, again indicating
good calibration. In Fig.
\ref{fig_blue_frac} we present the blue fraction ($f_b$) for the new clusters discovered optically
(filled circles) compared to the CNOC clusters (open circles). Cluster \#3353 does not have the
multi-color data necessary to compute a blue fraction, and is omitted. We have also plotted field blue
fractions calculated from the CNOC field galaxy data (crosses without errors), with color cuts
calculated assuming the same redshifts and magnitude limits as each of the 8 new clusters. These points
illustrate the values of $f_b$ we would expect for each of our new clusters if they had purely field
galaxy populations. We see that although a few objects have high values of
$f_b$ (notably
\#2906 at $f_b \approx 0.86$, one of our possible cluster identifications; see Figure \ref{fig_2906}),
as a sample, the new clusters exhibit a similar range of
$f_b$ with redshift as the CNOC clusters, suggesting that they are not in a significantly different
evolutionary state, at least as far as galaxy populations are concerned. For all the clusters
 except \#2906,
the blue fractions are well below the corresponding CNOC field values, indicating that we have clearly
detected an overdensity of red galaxies in each case.
\begin{\myfigure}
\epsscale{\figscale}
\plotone{\figdirthree 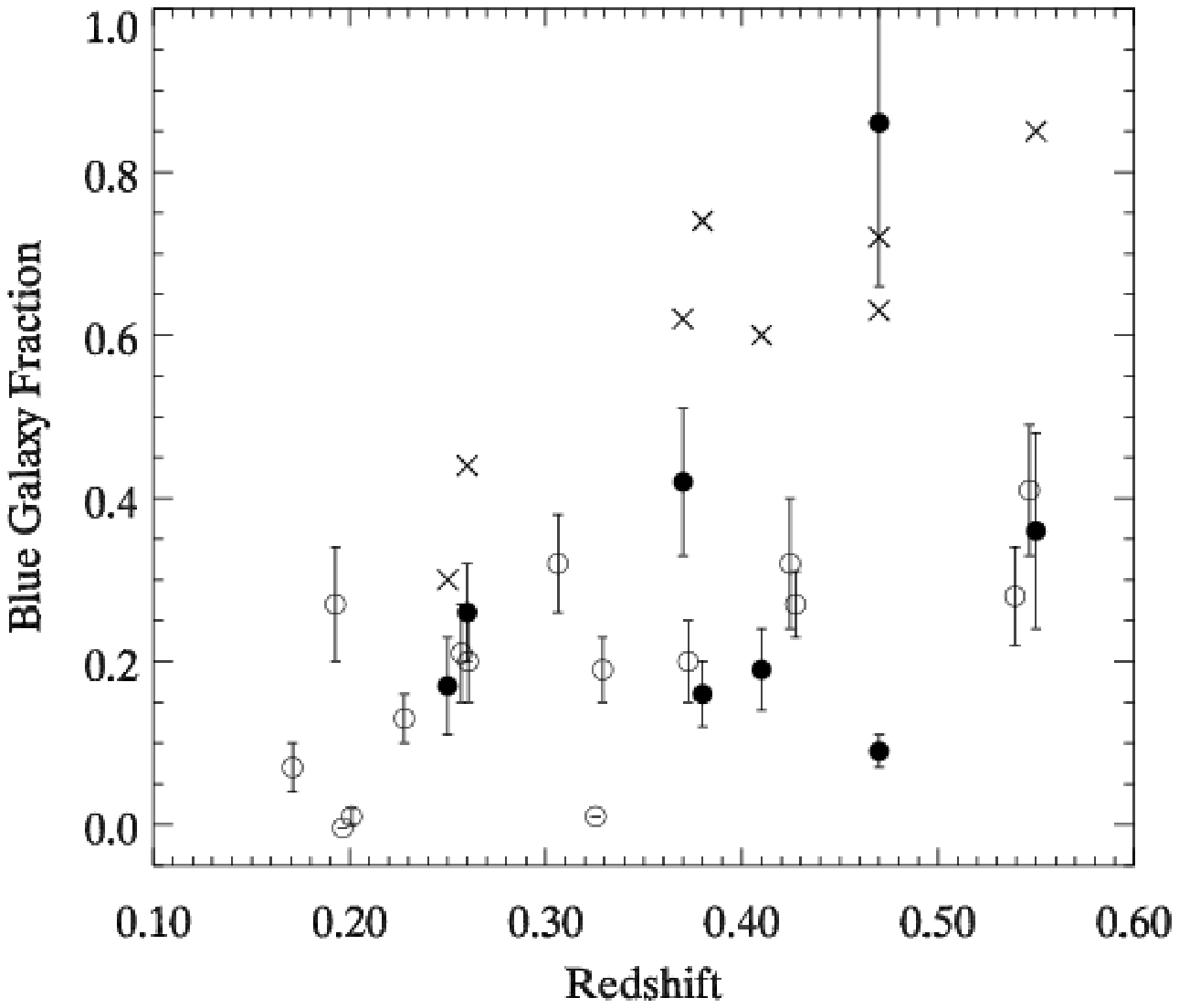}
\caption[Blue galaxy Fraction vs. Redshift]{Blue galaxy fraction vs. redshift for the optically
discovered new clusters (filled circles), and the CNOC clusters (open circles). Crosses indicate the
field galaxy blue fractions (used as the background subtraction) for each of the new cluster fields.}
\label{fig_blue_frac}
\end{\myfigure}
\epsscale{1.0}

To make certain that the methodology we have used to compute $f_b$ is reasonable, we
have compared our photometric $f_b$ values for the CNOC clusters with previous work.
\citet{ell01} and
\citet{kod01} measured $f_b$ for several of the CNOC clusters using a spectroscopic and another
photometric method, respectively. 
The mean of the $f_b$ values from the three samples is not significantly different, while individual
cluster comparisons indicate only minor differences.

In summary, while these new clusters appear to have more diffuse or asymmetric X-ray emission,
our optical comparisons do not indicate that they are significantly different from other EMSS
clusters. The
verification that these new clusters have similar properties to previously discovered EMSS clusters
(excepting the possible cluster \#2906) also substantiates their identification as X-ray clusters.


\section{Conclusions\label{sec_conclusions3}}

We have obtained deep images and multi-color photometry for
several new, rich, X-ray luminous clusters of galaxies apparently
missing from the EMSS sample. We have estimated the redshift of
these clusters based on the detection of a red sequence of
galaxies combined with a significant overdensity of galaxies
relative to field galaxy counts measured by the B$_{gc}$
statistic. We have presented the images and color-magnitude (CM)
diagrams of these clusters, with a detailed discussion of each
object, to elucidate our identification criteria. We have shown
that the X-ray and optical properties of these clusters ($L_X$,
B$_{gc}$, and the fraction of blue galaxies, $f_b$)  are
consistent with those expected of massive clusters at our
photometric redshift estimates. We have compared the photometric
optical properties of the new clusters with the CNOC clusters,
showing that as a sample, the new clusters (excepting one
``possible'' cluster ID: \# 2906)  exhibit no significant
differences in either galaxy overdensity (B$_{gc}$) or galaxy
populations ($f_b$). Thus we have confirmed the existence and
reality of these sources as clusters of galaxies and so validate
the analysis of Paper 1, which adds these new clusters to the EMSS
complete flux-limited cluster sample. We found in Paper 1 that when these clusters are added to the
X-ray luminosity function, the evidence for ``negative'' evolution in X-ray
luminous clusters out to
$z\sim$0.5 has been reduced to $\leq1\sigma$.

From our X-ray observations, we have presented evidence that many of these
clusters are more diffuse than typical EMSS clusters at similar
redshifts using flux ratios obtained through different size
apertures (see \S \ref{sec_discussionp3}, and Fig. \ref{fig_apcompare}).  However, the IPC data used for
this comparison is of moderate resolution and often poor S/N (see Paper 1), preventing
us from analyzing surface brightness profiles for these clusters
in detail. So we do not know at
present whether these new clusters are examples of true LSB
clusters or rather have only moderately asymmetric or ``lumpy''
X-ray morphologies, which prevented their detection by the
original EMSS detect cell size and shape. In either case
these new clusters have been selected
against by the EMSS survey and perhaps could be selected against by other surveys as well 
\citep[see e.g.,][]{ada00}.

There are two possible outcomes of a detailed X-ray and dynamical investigation of these clusters: 

\noindent (1)
They are shown to be virialized, and only moderately asymmetrical. In this case, we expect them to have
high $T_X$ (and thus high masses) consistent with their $L_X$, and they will contribute significantly 
to the X-ray temperature function
\citep[see e.g.,][]{don99a,hen00}, in a way that will push current
estimates of $\Omega_{matter}$ lower. Likely, they will have galaxy dynamics similar to previously
well-studied EMSS/CNOC clusters (as already suggested by $f_b$), thus fitting into the current general
picture of X-ray selected cluster properties.

\noindent (2) These clusters will be identified as X-ray LSB objects. To have high overall luminosities
with diffuse emission, it is expected that the ICM will be of lower temperature. Such clusters cannot
simultaneously follow scaling relations between mass and
$T_X$ or $L_X$, and may be significantly departed from gravitational equilibrium. This is a plausible
description of a class of clusters in the process of formation. In this case, the similar galaxy
populations observed suggests that galaxy evolution (i.e., Butcher-Oemler effect) occurs primarily
outside of virialized environments, raising new questions about the formation of groups and clusters.
We will also have increased our knowledge of the range of properties of X-ray selected clusters.

We intend to pursue high resolution and S/N X-ray imaging
with {\it Chandra} or {\it XMM-Newton}, in order to detect or
exclude LSB X-ray morphology, as well as determine $T_X$, thus
measuring the depth of the potential well for these new EMSS clusters. Extensive optical
spectroscopy is planned that will provide galaxy velocity
dispersions, definitive cluster membership information and galaxy
spectral energy distributions. 
The next generation of X-ray cluster surveys are now underway with
\textit{XMM-Newton} and \textit{Chandra}, and will yield
unprecedented numbers of new clusters, and new insight into X-ray
cluster properties measured with greater precision than ever
before. We emphasize that our studies have detected
real additions to the EMSS sample, due
to details of their X-ray morphology (asymmetries and/or LSB). 
A failure to properly account for clusters such as these could bias
the results of future studies. A detailed analysis determining the
prevalence of diffuse clusters within a flux-limited, X-ray selected
sample will increase the robustness of such a survey's
cosmological constraints.

\myacknowledgments

The authors wish to thank all those who shared data and information with us, including Isabella
Gioia, Tesla Jeltama, Tadayuki Kodama, Eric Perlman, Kathy Romer, and Howard Yee. ADL wishes to thank
the KPNO support staff during the observing runs for this project, Michael Harvanek for help with data
analysis and reduction, Ka Chun Yu for help with figures, the HEASARC facilities for maintaining
invaluable research tools and databases, Howard Yee for assistance and support using PPP, and Beth
White for her ongoing support. ADL, JTS, and this work were supported by a NASA Astrophysical Data
Program grant
\#NAG5-6936 and by travel grants for thesis work at Kitt Peak National Observatory by NOAO. EE
acknowledges support provided by the National Science Foundation grant AST 9617145. This work has made
use of the US Naval Observatory's Finder Chart Service available from
\url{http://www.nofs.navy.mil/data/FchPix} \citep{lev00}.

\begin{deluxetable}{lllclcll}
\tabletypesize{\footnotesize}
\tablecaption{Summary of Optical Observations\label{tab_obs_data}}
\tablewidth{0pt}
\tablehead{
\colhead{Cat. \#} & \colhead{R.A. (J2000)} & \colhead{decl. (J2000)} & \colhead{Filter} & \colhead{Exp. (s)}
& \colhead{$m_{lim}$} & \colhead{UT Date} & \colhead{Comments}
}
\startdata
161  & 00	53	10.38	&	+01	52	25.3	& $r$ & 1800 & 23.0 & 19 Dec 1998 & Cirrus \\
& & & $B$ & 600 & \nodata & 19 Dec 1998 & Cirrus \\
& & & $r$ & 600 & 22.8 & 22 Dec 1998 & Photometric \\
& & & $i$ & 1350 & 22.6 & 22 Dec 1998 & Photometric \\
& & & $B$ & 600 & \nodata & 22 Dec 1998 & Photometric \\
& & & $g$ & 2700 & 24.1 & 23 Dec 1998 & Photometric \\
1310 & 07 16 39.58 & +37 19 27.8 & $r$ & 2400 & 23.6 & 19 Dec 1998 & Cirrus \\
& & & $B$ & 600 & \nodata & 19 Dec 1998 & Cirrus \\
& & & $r$ & 600 & \nodata & 20 Dec 1998 & Photometric \\
& & & $B$ & 600 & \nodata & 20 Dec 1998 & Photometric \\
& & & $i$ & 1350 & 22.7 & 20 Dec 1998 & Photometric \\
& & & $g$ & 2700 & 24.6 & 23 Dec 1998 & Photometric \\
& & & $r$ & 600 & \nodata & 23 Dec 1998 & Photometric \\
1492 & 08 51 40.06 & +33 31 23.3 & $B$ & 600 & \nodata & 24 Dec 1998 & Photometric \\
& & & $r$ & 1800 & 23.6 & 24 Dec 1998 & Photometric \\
& & & $g$ & 2700 & 24.3 & 24 Dec 1998 & Photometric \\
& & & $i$ & 1200 & 23.1 & 24 Dec 1998 & Photometric \\
1605 & 09 39 58.76 & $-$02 49 26.6 & $r$ & 1800 & 23.6 & 24 Dec 1998 & Photometric \\
& & & $B$ & 600 & \nodata & 24 Dec 1998 & Photometric \\
& & & $g$ & 2700 & 24.2 & 24 Dec 1998 & Photometric \\
1681  & 10 07 51.41 & +12 45 43.6 & $R$ & 1800 & 24.0 & 04 May 2000 & Photometric \\
& & & $I$ & 1200 & 22.8 & 04 May 2000 & Photometric \\
& & & $B$ & 450 & \nodata & 04 May 2000 & Photometric \\
2436 & 13 30 11.38 & +30 43 51.3 & $R$ & 2400 & 24.2 & 03 May 2000 & Photometric \\
& & & $V$ & 1800 & 24.4 & 03 May 2000 & Photometric \\
& & & $I$ & 1200 & 22.8 & 03 May 2000 & Photometric \\
& & & $B$ & 450 & \nodata & 03 May 2000 & Photometric \\
2844 & 15 12 27.90 & +72 00 47.6 & $R$ & 1800 & 24.0 & 01 May 2000 & Variable Seeing \\
& & & $B$ & 450 & \nodata & 01 May 2000 & Variable Seeing \\
& & & $I$ & 1200 & 22.4 & 01 May 2000 & Variable Seeing \\
& & & $V$ & 1800 & 24.0 & 02 May 2000 & Variable Seeing \\
2906 & 15 31 55.38 & +24 20 42.9 & $R$ & 1800 & 24.3 & 04 May 2000 & Photometric \\
& & & $I$ & 1350 & 22.8 & 04 May 2000 & Photometric \\
& & & $B$ & 300 & \nodata & 04 May 2000 & Photometric \\
3353 & 17 55 51.73 & +67 53 44.9 & $R$ & 1800 & \nodata & 01 May 2000 & Photometric \\
& & & $V$ & 1350 & \nodata & 01 May 2000 & Photometric \\
\enddata
\tablecomments{$m_{lim}$ entries were omitted for observations which were not used in the galaxy color or richness
analyses.}
\end{deluxetable}
\begin{deluxetable}{lcccccl}
\rotate
\tabletypesize{\small}
\tablecaption{New Clusters in the EMSS\label{tab_newclusters3}}
\tablewidth{0pt}
\tablehead{
\colhead{Cat. \#}  & \colhead{$z$} & \colhead{B$_{gc}$} & \colhead{log $L_X$\tablenotemark{a}} &
\colhead{$f_b$\tablenotemark{b}} & \colhead{Ap3/Ap1\tablenotemark{c}} & \colhead{Notes} \\ 
& & \colhead{[\bggmph]} & & & Ratio & \\
}
\startdata
161$^\dagger$  & $0.52-0.59$ [0.55] & $610-2230$ [1340] & $44.7-45.0$ [44.8] & $0.36\pm0.11$ & 3.75 & Dominant X-ray Source \\   
&	$0.32-0.38$ [0.35] & $500-980$ [740] & $44.3-44.5$ [44.4] & \nodata & \nodata & 2nd Cluster in Field \\
1310 & $0.34-0.40$ [0.37] & $790-1410$ [1200] & $44.5-44.9$ [44.7] & $0.42\pm0.09$  & 3.57 & \\
1492 & $0.42-0.50$ [0.47] & $1550-2910$ [2240] & $44.9-45.3$ [45.1] & $0.09\pm0.04$ & 1.78 & \\
1605 & $0.22-0.27$ [0.25] & $670-1050$ [830] & $43.9-44.2$ [44.1] & $0.17\pm0.07$ & 1.86 & \\
1681 & $0.35-0.45$ [0.41] & $180-1030$ [600] & $44.1-44.6$ [44.4] & $0.19\pm0.10$ & 1.67 & \\
2436 & $0.22-0.28$ [0.26] & $1120-1680$ [1450] & $44.1-44.4$ [44.3] & $0.26\pm0.05$ & 2.33 & GHO Cluster, Dominant X-ray Source \\ 
& $0.27-0.33$ [0.30] & $400-650$ [520] & \nodata & \nodata & \nodata & 2nd Cluster 5\arcmin~NE \\
2844$^\dagger$ & $0.33-0.44$ [0.38] & $640-1260$ [820] & $44.3-44.8$ [44.6] & $0.16\pm0.09$ & 3.00 & Variable Seeing \\
2906$^\dagger$ & $0.42-0.53$ [0.47] & $390-1410$ [970] & $44.6-45.0$ [44.8] & $0.86\pm0.08$ & 1.27 & BCG 2\arcmin~NE, Possible
Contamination \\
3353 & $<0.1$ & \nodata & $<44.0$ & \nodata & \nodata & Nearby Group; No B$_{gc}$ Measured \\
\enddata
\tablecomments{Best fit values for redshift, B$_{gc}$, and $L_X$ are given in [square brackets] after the range. Clusters denoted
with a
$^\dagger$ are possible cluster identifications, pending further investigation.}
\tablenotetext{a}{Log of the cluster X-ray luminosity in the \eband~band in \lxh. Employing alternative cosmologies
(e.g., H$_0=70$~km~s$^{-1}$~Mpc$^{-1}$, $\Omega_0=0.3$, $\Lambda=0.7$) reduces these $L_X$ values (a maximum of 30\% at $z=0.5$). IPC
luminosity is calculated from the third aperture count rate, see Paper 1; the range of values includes an estimate of the Poisson error
on the detected count rate.}
\tablenotetext{b}{Photometrically defined blue galaxy fraction, as shown in Fig. \ref{fig_blue_frac}. Errors are based on the numbers of
galaxies detected within the color cuts.}
\tablenotetext{c}{Ratio of IPC aperture 3 to aperture 1 flux, indicating X-ray extent (see \S \ref{sec_discussionp3}).}
\end{deluxetable}

\end{document}